\magnification \magstep1
\input amssym.def
\input amssym.tex

\bigskip
\bigskip
\bigskip
\bigskip
\bigskip
\centerline{\bf Quantum Amplitudes in Black-Hole Evaporation:}
\centerline{\bf Coherent and Squeezed States} 
\bigskip
\centerline{A.N.St.J.Farley and P.D.D'Eath}
\bigskip
\bigskip
\centerline{Department of Applied Mathematics and Theoretical Physics,
Centre for Mathematical Sciences,} 
\smallskip
\centerline{University of Cambridge, Wilberforce Road, 
Cambridge CB3 0WA, United Kingdom} 
\bigskip
\centerline{Abstract}
\bigskip
In earlier papers, 
the quantum amplitude for processes involving the formation 
and evaporation of black holes
was calculated by means of a complex-time approach. 
Instead of taking a more familiar approach to black-hole evaporation, 
we simply followed Feynman's 
$+{\,}i\varepsilon$ 
approach in quantum field theory. 
The Lorentzian time-interval 
${\,}T{\,}$, 
measured at spatial infinity between a pair 
of asymptotically-flat space-like hypersurfaces
$\Sigma_{I}$ 
and 
$\Sigma_{F}$ 
carrying initial and final boundary data 
for the gravitational and other fields, 
is rotated: 
$T\rightarrow{\mid}T{\mid}{\,}\exp(-{\,}i\delta){\,}$, 
where 
${\,}0{\,}<{\,}\delta{\,}\leq{\pi}/2{\,}$. 
Classically and quantum-mechanically, 
this procedure is expected to lead to a well-posed boundary-value problem. 
Thus, 
what we have done is to find quantum amplitudes 
(not just probability densities) 
relating to a pure state at late times 
following gravitational collapse of matter to a black hole. 
Such pure states, 
arising from gravitational collapse, 
are then shown to admit a description in terms 
of coherent and squeezed states. 
Indeed, 
this description 
is not so different from that arising in a well-known context, 
namely, 
the highly-squeezed final state of the relic radiation background 
in inflationary cosmology. 
For definiteness, 
we study the simplest model of collapse, 
based on Einstein gravity with a massless scalar field. 
Following the complex rotation above, 
one finds that, 
in an adiabatic approximation, 
the resulting quantum amplitude may be expressed in terms 
of generalised coherent states of the harmonic oscillator. 
A physical interpretation is given; 
further, 
a squeezed-state representation follows.\par

\medskip

\noindent {\bf 1. Introduction }
\medskip

\noindent 
The treatment given here of black-hole evaporation 
depends on Feynman's 
$+{\,}i\varepsilon$ 
approach [1].
We begin by reviewing this approach, 
which in [2-8] was described and applied to quantum amplitudes 
(not just probabilities) 
for particle production, 
following gravitational collapse to a black hole. 
In specifying either the classical boundary-value problem 
or the quantum amplitude to be computed, 
one takes an initial space-like ypersurface 
$\Sigma_{I}$ 
and a final space-like hypersurface 
$\Sigma_{F}{\,}$. 
For simplicity, 
imagine that we are in the asymptotically-flat context and, 
further, 
that each of 
$\Sigma_{I}$ 
and 
$\Sigma_{F}$ 
has the topology of 
$\Bbb R^{3}{\,}$. 
One then specifies suitable asymptotically-flat boundary data 
for Einstein gravity and for any other fields present on 
$\Sigma_{I}$ 
and 
$\Sigma_{F}{\,}$. 
If there are no fermionic fields present in the Lagrangian, 
then one might expect the gravitational boundary data 
to consist of the intrinsic positive-definite 
(Riemannian) 
3-geometry, 
given by the intrinsic 3-metric 
${\,}h_{ij}=g_{ij}{\,}$ 
$(i,j=1,2,3)$ 
on a surface
$\{x^{0}={\rm const. }\}{\,}$.  
Here, 
${\,}g_{\mu\nu}$ 
${\,}(\mu,\nu=0,1,2,3)$ 
gives the 4-dimensional `space-time' metric.
For a complete specification of the problem 
(if indeed such a boundary-value problem is well-posed{~} -- see below), 
one further needs to give the Lorentzian proper-time interval 
$T$
which separates 
$\Sigma_{I}$ 
from 
$\Sigma_{F}{\,}$, 
as measured at spatial infinity.  
In the papers [3,4], 
as in the present paper, 
we take the simplest possibility for the matter fields present, 
namely, 
a real massless scalar field 
$\phi{\,}$.  
Then, 
suitable boundary data on 
$\Sigma_{I}$ 
and on 
$\Sigma_{F}$ 
are expected to be 
${\,}(h_{ij},\phi)_{I,F}{\,}$.  
For further simplicity, 
without great conceptual loss, 
we assume here that the initial data on 
$\Sigma_{I}{\,}$ 
are spherically symmetric, 
corresponding to a diffuse slowly-moving initial matter distribution, 
prior to gravitational collapse.
The final gravitational and scalar data 
are taken to have a `background' spherically-symmetric part, 
together with small non-spherical perturbations, 
which correspond to gravitons and massless-scalar particles; 
such data represent a possible field configuration 
on a space-like slice of topology 
$\Bbb R^{3}$ 
at late times, 
for which a quantum amplitude should be calculable.

In Feynman's 
$+{\,}i\varepsilon$ 
approach [1], 
one rotates the time-interval 
$T$ 
into the lower-half complex plane:
$T\rightarrow{\mid}T{\mid}{\,}\exp(-{\,}i\delta){\,}$, 
where 
${\,}0{\,}<{\,}\delta{\,}\leq\pi/2{\,}$. 
Except in the extreme case 
$\delta{\,}={\,}\pi/2{\,}$, 
one has classically a complex boundary-value problem, 
with a complex classical 4-metric 
${\,}g_{\mu\nu}$ 
and scalar field 
${\,}\phi{\,}$ 
in the interior, 
for the given 
(real) 
boundary data 
$(h_{ij}{\,},\phi)$ 
on 
${\,}\Sigma_{I}$ 
and 
${\,}\Sigma_{F}{\,}$. 
In the special case 
${\,}\delta{\,}={\,}\pi/2{\,}$, 
one has a real Riemannian time separation 
${\,}|T|$ 
at spatial infinity, 
and therefore one expects to have a well-behaved real elliptic 
boundary-value problem 
(${\it modulo}$ gauge) 
for 
$(g_{\mu\nu}{\,},\phi){\,}$. 
From comparison with the case 
of linearised massless scalar fields [4,9], 
one expects that the complex case 
$(0{\,}<{\,}\delta{\,}<\pi/2)$ 
should be 
${\it strongly{\,} elliptic}$ 
[10]; 
that is, 
despite the complex boundary parameter 
$T{\,}$, 
the boundary-value problem should continue to be well-posed, 
with analytic solutions, 
existence and uniqueness. 
By contrast, 
the `boundary-value problem' for
${\,}\delta{\,}={\,}0{\,}$, 
in which one asks for a real Lorentzian-signature solution 
to the boundary-value problem, 
is well known to be ill-posed [11]: 
it is inappropriate to pose boundary-value problems for hyperbolic 
(wave-like) 
systems.\par
\smallskip
\indent 
For the black-hole collapse problem, 
we consider boundary data of the type described above, 
with weak-field linear-order anisotropic perturbations 
$\bigl(\lambda{\,}h^{(1)}_{ij}{\,},
{\,}\lambda{\,}\phi^{(1)}\bigr)$ 
away from spherical symmetry on 
$\Sigma_{F}{\,}$, 
but with no anisotropy on 
$\Sigma_{I}{\,}$, 
where 
$\lambda$ 
is a small parameter. 
The Lorentzian classical action of the exactly spherically-symmetric solution 
(here assuming ${\,}\delta{\,}>0{\,}$) 
described by 
${\,}g_{\mu\nu}^{(0)}{\,}={\,}\gamma_{\mu\nu}{\,}$
and 
${\,}\phi^{(0)}{\,}={\,}\Phi{\,}$, 
will be written as 
$S_{\rm class}^{(0)}{\,}$. 
Given the small 
(but non-zero) 
anisotropy of the final data and hence of the interior classical solution 
$(g_{\mu\nu}{\,},{\,}\phi){\,}$, 
one considers the asymptotic expansions:
$$\eqalign{g_{\mu\nu}{\quad}
&\sim{\quad}\gamma_{\mu\nu}{\;}{\,}
+{\;}{\,}\lambda{\,}g_{\mu\nu}^{(1)}{\;}{\,}
+{\;}{\,}\lambda^{2}{\,}g_{\mu\nu}^{(2)}{\;}{\,}
+{\;}{\,}{\ldots}{\quad},\cr
\phi{\quad}
&\sim{\quad}\Phi{\;}{\,}
+{\;}{\,}\lambda{\,}\phi^{(1)}{\;}{\,}
+{\;}{\,}\lambda^{2}{\,}\phi^{(2)}{\;}{\,}
+{\;}{\,}{\ldots}
{\quad}.\cr}\eqno(1.1)$$
For such linearised boundary data, 
with corresponding perturbative expansions (1.1) 
for the classical solution, 
the classical Lorentzian action 
${\,}S_{\rm class}{\,}$ 
will have an asymptotic expansion of the form:
$$S_{\rm class}{\quad}
\sim{\quad}S_{\rm class}^{(0)}{\;}{\,}
+{\;}{\,}S_{\rm class}^{(2)}{\;}{\,}
+{\;}{\,}S_{\rm class}^{(3)}{\;}{\,}
+{\;}{\,}{\ldots}
{\quad}.\eqno(1.2)$$ 
Here, 
the perturbative factors 
${\,}\lambda^{2}$ 
multiplying 
$S_{\rm class}^{(2)}{\;}$, 
${\,}\lambda^{3}$ 
multiplying 
${\,}S_{\rm class}^{(3)}{\;}$, 
etc., 
have been omitted, 
in order to simplify the notation later.
The first correction 
$S_{\rm class}^{(2)}$ 
is the second-variation classical action and is bilinear in the linear-order 
${\,}\bigl(\lambda{\,}h^{(1)}_{ij}{\,},
{\,}\lambda{\,}\phi^{(1)}\bigr)$ 
corrections to the boundary data on 
$\Sigma_{F}{\,}$. 
As in [7], 
one can indeed evaluate 
${\,}S_{\rm class}^{(2)}{\,}$ 
jointly as a functional of the linearised boundary data 
${\,}\bigl(\lambda{\,}h^{(1)}_{ij}{\,},
{\,}\lambda{\,}\phi^{(1)}\bigr)_{F}{\,}$ 
and a function of the complex variable 
$T{\,}$. 
One then computes the corresponding semi-classical quantum amplitude, 
proportional to 
${\,}\exp\bigl(iS_{\rm class}^{(2)}\bigr){\,}$, 
and one can also include loop corrections, 
if appropriate. 
In fact, 
it appears likely, 
first, 
that any quantum field theory which includes gravity 
must be invariant under local supersymmetry
(and hence a theory involving supergravity) 
if it is to have meaningful quantum amplitudes, 
and, 
second, 
that such a quantum theory may actually have finite quantum amplitudes, 
even though these may include loop corrections, 
unless 
(for example) 
the theory is pure 
$N=1$ 
supergravity without supermatter [9,12]. 
In that case, 
the quantum amplitudes considered in the black-hole problem 
can certainly be truncated as 
${\,}\exp\bigl(iS_{\rm class}^{(2)}\bigr)$, 
provided that the frequencies involved in the boundary data
$\bigl(\lambda{\,}h^{(1)}_{ij}{\,},
{\,}\lambda{\,}\phi^{(1)}\bigr)$ 
are well below the Planck frequency. 
Finally, 
then, 
following Feynman's 
$+{\,}i\varepsilon$ 
prescription in the present context of black-hole quantum evaporation, 
one recovers the Lorentzian quantum amplitude 
(again, not just the probability density) 
for the quantum state including 
(say) 
created particles present at late times, 
by taking the limit of the semi-classical amplitude 
${\,}\exp\bigl(iS_{\rm class}^{(2)}\bigr)$ 
as 
${\,}\delta{\,}\rightarrow{\,}0_{+}{\,}$.
\smallskip
\indent 
As seen in [4,6,8],
the black-hole radiation has the usual thermal spectrum,
at the temperature
${\,}1/8{\pi}M_{I}{\;}$.
Unless otherwise stated,
we employ Planckian units,
taking:
${\,}k_{B}{\,}=c{\,}={\,}\hbar{\,}={\,}G{\,}={\,}1{\,}$.
Here,
${\,}M_{I}{\,}$
is the ADM
(Arnowitt-Deser-Misner)
mass of the 'space-time' [27].
As described in [3] and References therein,
the ADM mass on the initial surface
$\Sigma_{I}{\,}$
must equal the ADM mass on the final surface
$\Sigma_{F}{\;}$,
in order that the classical boundary-value problem should be well-posed,
and in particular that the space-time metric components 
should have the expected fall-off properties at spatial infinity.
In the slightly-complexified r\'egime,
where the
(large)
time-interval
${\,}T{\,}$
at spatial infinity obeys
${\,}{\rm Im}(T){\,}<{\,}0{\,}$,
the geometry is accurately approximated by a Vaidya metric [6,20],
with slowly-varying mass function
${\,}m(u){\,}$,
where
${\,}u{\,}$
is a 'retarded-time' coordinate.
If one wished,
one could pursue the
(classical)
perturbation theory of [6] further,
so as to describe accurately the slightly-complex classical geometry,
given anisotropic weak-field final data on the late-time hypersurface
$\Sigma_{F}{\;}$,
whose mass agrees,
as above,
with the mass of the initial-data set on
$\Sigma_{I}{\;}$.
That is,
not only would one arrive at the usual retarded-time dependence of
$m(u){\,}$
for a radiating black hole{~}
-- {~}one would also
(given sufficient labour)
calculate the detailed evolution backwards in time
of the weak-field anisotropic perturbations.\par
\smallskip
\indent
Of course,
the usual definition of a black hole depends on the space-time metric
being real,
of Lorentzian signature.
In contrast,
the classical Einstein boundary-value problem 
is only expected to be well-behaved when
${\,}{\rm Im}(T){\,}<{\,}0{\,}$,
for a time-interval
${\,}T{\,}$
measured at infinity.
Thus,
strictly speaking,
it is inappropriate to use the term 'black hole' in relation
to the complexified classical solution of the previous paragraph.
However,
it would be a fair use of language 
to say that the infilling Vaidya-like classical solution
for the boundary-value problem of the previous paragraph
is a 'black-hole intermediate state'.\par
\smallskip
\indent
Readers who are accustomed to the original Lorentzian-signature
derivation of black-hole evaporation will be used to the notion
of radiation
(scalar-field, Mazwell, etc.)
piling up around the future event horizon,
and then undergoing an enormous redshift,
depending in a specific way on the mass
$M_{I}{\;}$,
as the radiative fields move out towards future null infinity.
The detailed form of this redshift 
is intimately connected with the thermality and temperature 
of the black-hole radiation.
But this radiative behaviour 
can be learnt equally well through study of high-frequency or 
$WKB$
solutions of the separated wave equation,
as in Eqs.(2.4,5) below.
The 
$WKB$
transmission and reflection coefficients describe the thermal radiation.
And
$WKB$
investigation of wave equations such as Eqs.(2.4,5)
is also at the base of the present approach [4,6,8].
One might say that detailed knowledge concerning the 
(Lorentzian)
event horizon is 'imbedded' in the relevant spin-$s$
wave equations,
such as Eqs.(2.4,5).
Thus,
it should not be surprising that one can arrive at the thermal spectrum
using either geometrical considerations of the Lorentzian horizon
or analytic arguments requiring a complex metric
(for which case the event horizon is undefined).
In similar fashion,
one may still speak here,
if desired,
about particles which fall into the hole as well as particles
which travel out to infinity;
these correspond to the usual basis of two
$WKB$
solutions of the radial wave equation.\par
\smallskip
\indent
In the present paper, 
we study quantum amplitudes found 
{\it via}
Feynman's approach, 
as discussed and calculated in [2-8], 
but now in the context of coherent states [13], 
which resemble `classical states', 
and of squeezed states [14], 
which are purely quantum-mechanical. 
As above, 
our motivation originated with the study of the final radiation 
which remains after a black hole has evaporated completely, 
but there are strong connections 
also with the relic Cosmic Microwave Background Radiation (CMBR) 
induced by inflationary cosmological perturbations. 
In fact, 
particle creation by black holes has several similarities 
with cosmological particle creation, 
despite the lack of asymptotic flatness in the cosmological case. 
Cosmological and black-hole particle creation 
both require a time-dependence in the metric. 
This in turn has led to descriptions, 
in terms of coherent and squeezed states, 
of quantum phenomena in curved space-time; 
some of the earlier examples include [15-17].\par
\smallskip
\indent
In inflationary cosmology, 
the field modes are in their adiabatic ground state, 
with short wavelengths near the start of inflation. 
This is related to the assumption that the universe 
was in a maximally-symmetric state at some time in the past [18,19]. 
Due to the accelerated expansion of the universe during inflation, 
quantum fluctuations 
are amplified into macroscopic or classical perturbations. 
The early-time fluctuations lead to the formation 
of large-scale structure in the universe, 
and also contribute to the anisotropies in the CMBR.  
The final state for the perturbations 
is a two-mode highly-squeezed state for modes whose radii 
are much greater than the Hubble radius, 
with pairs of field quanta 
(having opposite momenta) 
being produced at late times [16,17].
Tensor 
($s=2$) 
fluctuations in the metric, 
for example, 
are predicted to give rise to relic gravitational waves.
On the other hand, 
electromagnetic waves 
($s=1$) 
cannot be squeezed in the same way during the cosmological expansion, 
because they do not interact with the external gravitational field 
in the same way.\par
\smallskip
\indent 
In either the cosmological or the black-hole case, 
one works within an adiabatic approximation for the perturbative modes. 
Writing 
$k$ 
for a typical perturbative frequency, 
one requires 
${\,}k{\;}{\gg}{\;}aH{\,}$ 
in the cosmological case, 
where 
$H=(\dot a/a){\,}$, 
with 
${\,}a(t)$ 
denoting the scale factor.  
In the black-hole case, 
the space-time geometry at late times, 
in the region containing a stream of outgoing radiation, 
is given by a Vaidya metric [20-21] 
with a slowly-varying `mass function' 
${\,}m(t,r){\,}$.  
The adiabatic condition then reads 
${\,}k{\;}{\gg}{\;}{\mid}{\dot m}{\mid}/m{\,}$. 
Indeed, 
for an evaporating black hole, 
at all except the last moments of evaporation, 
the frequencies of interest in the evaporating modes 
do typically exceed 
${\mid}{\dot m}{\mid}/m{\,}$, 
namely, 
the inverse time-variation scale for the black-hole mass. 
In other words, 
the period of a wave of interest 
is typically much smaller than the time-scale for variations 
in the background gravitational field. 
The black hole interacts negligibly with the emitted particles, 
and the time between successive emissions 
is comparable with the black-hole mass [22].\par
\smallskip
\indent 
In the cosmological case, 
a physical description of the corresponding phenomenon is that, 
when the wavelength is comparable with or larger than the Hubble radius, 
amplification of the zero-point quantum fluctuations takes place. 
As a further aspect of this analogy, 
the redshifting of the radiation in the black-hole background space-time 
is determined by the total mass 
$M_{I}{\,}$, 
and correspondingly by the Hubble parameter 
$H^{-1}$ 
in the cosmological case.\par
\smallskip
\indent 
In this paper, 
we apply the squeezed-state formalism to black-hole evaporation, 
while maintaining a comparison with inflationary cosmology 
-- see [23] for further comparisons.  
In the case of inflationary cosmology, 
the quantum evolution of cosmological perturbations 
(density, rotational and gravitational), 
which begin in an initial vacuum state, 
follows essentially a set of Schr\"odinger equations [19]. 
The state of the perturbations 
is transformed into a highly-squeezed vacuum state, 
with many particles, 
having a large variance in their amplitude 
(particle number), 
but small 
(squeezed) 
phase variations. 
The squeezing of cosmological perturbations 
may be suppressed at small wavelengths, 
but it should be present at long wavelengths, 
especially for gravitational waves [24]. 
These perturbations also induce the anisotropies at large angular scales, 
as observed in the CMBR. 
Their wavelengths today 
are comparable with or greater than the Hubble radius. 
The above amplification of the initial zero-point fluctuations 
gives rise to standing waves with a fixed phase, 
rather than travelling waves. 
The relic perturbations in the high-squeezing or WKB limit 
can be described as a stochastic collection of standing waves [16,17]. 
This paragraph has reviewed the cosmological case; 
as will be seen below, 
a similar picture emerges in the application 
to black-hole evaporation.\par
\smallskip
\indent 
In Sec.2, 
we outline the main features of the above complex approach 
to the calculation of quantum amplitudes 
(not just probabilities) 
for perturbative data 
(spins $s=0,1,2$) 
prescribed on a late-time final hypersurface 
$\Sigma_{F}{\,}$. 
For this procedure to be well-posed, 
one has first to rotate:
$T\rightarrow{\mid}T{\mid}{\,}\exp(-{\,}i\delta)$ 
into the lower half-plane. 
The resulting standing waves, 
which originate from setting Dirichlet boundary data 
on the initial and final space-like hypersurfaces, 
turn out to correspond to a highly-squeezed final state 
for late-time black-hole radiation. 
In the adiabatic approximation, 
the fixed phases correspond to discrete frequencies 
in the remnant 
(quantum) 
radiation from the evanescent black hole.\par
\smallskip
\indent 
Secs.3, 4, 5 describe coherent states, 
generalised coherent states 
and squeezed states, 
respectively. 
In Sec.6, 
the small angle 
$\delta{\,}$, 
through which the time 
$T$ 
at spatial infinity is rotated into the complex:
$T\rightarrow{\mid}T{\mid}{\,}\exp(-{\,}i\delta){\,}$, 
is related to the large amount of squeezing 
which has been applied to give the final state. 
We also discuss briefly the normalisation of the probability density, 
and demonstrate the existence of large peaks and troughs 
in the spectrum of the radiation reaching future null infinity, 
due to the standing-wave pattern of the perturbations. 
A short discussion of entropy and squeezing is given in Sec.7, 
and possible classical predictions are considered in Sec.8.  
Sec.9 contains a brief Conclusion.
Some technical results are given in the Appendix. 
A briefer account of this work has appeared in [25].\par

\medskip

\noindent {\bf 2. The quantum amplitude for late-time data}

\medskip

\indent 
Consider first the effect on the classical boundary-value problem 
of a rotation into the complex of the time-interval 
$T{\,}$ 
by a moderately small angle 
$\delta{\,}$, 
as above. 
The resulting classical solution 
$(g_{\mu\nu}{\,},{\,}\phi)$ 
of the coupled Einstein/massless-scalar field equations 
will be somewhat complexified, 
by comparison with a Lorentzian-signature solution.  
By suitable choice of coordinates 
${\,}(t,r,\theta,\varphi){\,}$,
the spherically-symmetric `background' part of the metric 
may be written in the form [2,4]
$$ds^{2}{\quad}
={\quad}-{\,}e^{b}{\;}dt^{2}{\,}
+{\;}e^{a}{\;}dr^{2}{\,} 
+{\,}r^{2}{\,}\bigl(d{\theta}^{2}
+{\,}{\sin}^{2}{\theta}{\;}d{\varphi}^{2}\bigr) 
{\quad},\eqno(2.1)$$ 
where 
${\,}b=b(t{\,},r){\,},
{\;}{\,}a=a(t{\,},r){\,}$, 
and the spherically-symmetric `background' part 
$\Phi$ 
of the scalar field has the form 
${\,}\Phi{\,}={\,}\Phi(t,r){\,}$. 
The spherically-symmetric functions 
$a{\,},{\,}b{\,}$ 
and 
$\Phi{\,}$ 
must, 
of course, 
be complex-valued. 
The coupled Lorentzian-signature Einstein/scalar field equations 
for this spherically-symmetric configuration 
are given by the analytic continuation of the Riemannian field equations 
Eqs.(1.9-13) of [7], 
on making the replacement
$$t{\quad}
={\quad}\tau{\,}\exp\bigl(-{\,}i\vartheta\bigr) 
{\quad},\eqno(2.2)$$ 
where 
$\tau$ 
is the `Riemannian time-coordinate' of [7], 
and where the real number 
$\vartheta$ 
should be rotated precisely from 
${\,}0{\,}$ 
to 
${\,}{\pi}/2{\,}$.\par
\smallskip
\indent 
The small anisotropic perturbations in the boundary data 
on the final late-time hypersurface 
${\Sigma}_{F}{\,}$ 
consist, 
in the language of Sec.1, 
of the perturbed part 
${\lambda}{\,}h_{ij}^{(1)}$ 
of the intrinsic 3-dimensional spatial metric 
$h_{ijF}$ 
on 
$\Sigma_{F}{\,}$, 
together with the perturbation 
${\lambda}{\,}{\phi}^{(1)}$ 
of the scalar field 
${\phi}_{F}$ 
on 
$\Sigma_{F}{\,}$. 
The classical solutions resulting from these perturbed boundary data 
correspond to gravitons and to massless-scalar particles, 
propagating on the 
(complex) 
spherically-symmetric classical background 
$\bigl(g_{\mu\nu}{\,},{\,}\Phi\bigr){\,}$. 
For example, 
the field 
${\,}\phi^{(1)}{\,}$ 
in the interior of the space-time may be decomposed as in Eq.(6) of [3]:
$$\phi^{(1)}(t,r,\theta,\varphi){\quad}
={\quad}{{1}\over{r}}{\;} 
\sum_{\ell =0}^{\infty}{\,}\sum_{m=-\ell}^{m=\ell}{\;} 
Y_{\ell m}(\Omega){\;}R_{\ell m}(t,r) 
{\quad}.\eqno(2.3)$$ 
Here, 
$Y_{\ell m}(\Omega){\,}$ 
denotes the 
${\,}(\ell,m)$ 
scalar spherical harmonic of [23].
The scalar field equation decouples for each 
$(\ell{\,},m){\,}$, 
leading to the mode equation
$$\Bigl(e^{(b-a)/2}{\,}{\partial}_{r}\Bigr)^{2}{\,}R_{\ell m}{\;} 
-{\,}\bigl(\partial_{t}\bigr)^{2}{\,}R_{\ell m}{\;}
-{\,}{{1}\over{2}}{\,}\Bigl(\partial_{t}(a-b)\Bigr){\,} 
\bigl(\partial_{t}R_{\ell m}\bigr){\,}
-{\;}V_{\ell}(t,r){\,}R_{\ell m}{\quad} 
={\quad}0 
{\;},\eqno(2.4)$$ 
where
$$V_{\ell}(t,r){\quad}
={\quad}{{e^{b(t,r)}}\over{r^{2}}}{\;} 
\biggl(\ell(\ell +1){\,}+{\,}{{2m(t,r)}\over{r}}\biggr) 
\eqno(2.5)$$
would be real and positive in the Lorentzian-signature case. 
The `mass function' 
$m(t,r){\,}$, 
which would equal the constant mass 
$M_{I}$ 
for an exact Schwarzschild geometry [27], 
is defined by
$$e^{-{\,}a(t,r)}{\quad}
={\quad}1{\,}-{\;}{{2m(t,r)}\over{r}} 
{\quad}.\eqno(2.6)$$ 
A corresponding angular decomposition 
can be given for weak gravitational-wave perturbations 
about the spherical background [7,28-30].\par
\smallskip
\indent 
In most regions of the classical space-time, 
except for the central region where the black hole is formed, 
the metric functions 
$a(t,r)$ 
and 
$b(t,r)$ 
must vary only `slowly' or `adiabatically'. 
This allows one to study, 
in such a region, 
a radial mode solution for 
(say) 
a perturbed scalar field, 
in which a further separation of the time dependence 
can be made approximately [2,25]:
$$R_{\ell m}(t,r){\quad}
\sim{\quad}\exp(ikt){\;}{\;}\xi_{k\ell m}(t,r) 
{\quad}.\eqno(2.7)$$ 
Here, 
${\,}\xi_{k\ell m}(t,r)$ 
varies `slowly' with respect to 
$t{\,}$.  
This will occur near spatial infinity and also, 
provided that the time-interval 
$T$ 
is sufficiently large, 
in a neighbourhood of the final hypersurface 
$\Sigma_{F}{\,}$.  
The mode equation (2.4,5) then reduces [4] to
$$e^{(b-a)/2}{\;}{{\partial}\over{\partial r}}
\Biggl(e^{(b-a)/2}{\;}{{\partial\xi_{k\ell m}}\over{\partial r}}\Biggr){\;}
+{\;}\Bigl(k^{2}{\,}-{\,}V_{\ell}\Bigr){\;}\xi_{k\ell m}{\quad} 
={\quad}0 
{\quad}.\eqno(2.8)$$ 
As seen in [2,6], 
the spherically-symmetric background metric in this region 
can be represented to high accuracy by a Vaidya metric, 
which describes the outflow of massless matter, 
which is spherically symmetric, 
on the average.
The principal condition for the validity 
of the adiabatic expansion is [6] that
$${\mid}k{\mid}{\quad}
{\gg}{\quad}{{{\mid}{\dot m}{\mid}}\over{m}} 
{\quad}.\eqno(2.9)$$ 
\indent 
In studying the behaviour of solutions of the radial mode equation (2.8), 
it is natural to define a generalisation 
$r^{*}$ 
of the standard Regge-Wheeler or `tortoise' coordinate 
${\,}r_{S}^{*}{\,}$ 
in the Schwarzschild geometry [27], 
by
$${{\partial}\over{\partial r^{*}}}{\quad}
={\quad}e^{(b-a)/2}{\;}{{\partial}\over{\partial r}} 
{\quad}.\eqno(2.10)$$ 
In the exact Schwarzschild metric, 
this gives 
$$r^{*}_{s}{\quad} 
={\quad}r{\;} 
+{\,}2{\,}M_{I}{\,}\log\biggl(\Bigl(r/2M_{I}\Bigl)-1\biggr)
{\quad}.\eqno(2.11)$$ 
The approximate 
(adiabatic) 
mode equation (2.8) then reads
$${{\partial^{2}\xi_{k\ell m}}\over{\partial r^{*2}}}{\;}
+{\;}\Bigl(k^{2}{\,}-{\,}V_{\ell}\Bigr){\,}\xi_{k\ell m}{\quad} 
={\quad}0 
{\quad}.\eqno(2.12)$$ 
\indent 
The procedure adopted in [2,6] involves choosing a convenient set 
of suitable radial functions 
${\,}\{\xi_{k\ell m}(r)\}{\,}$ 
on the final surface 
$\Sigma_{F}{\,}$, 
since it is here that the non-trivial boundary data are posed.
The mode equation (2.12) does not depend on the quantum number 
$m{\,}$, 
whence we may choose 
${\,}\xi_{k\ell m}(r){\;}={\;}\xi_{k\ell}(r){\,}$, 
independently of 
$m{\,}$. 
The boundary condition of regularity at the spatial origin 
$\{r=0\}{\;}$ 
[4,6] implies that
$$\xi_{k\ell}(r){\quad}
={\quad}{\rm constant}{\,}{\times}{\,}\bigl(kr\bigr)^{\ell +1}{\;} 
+{\;}O\Bigl(\bigl(kr\bigr)^{\ell +3}\Bigr)
{\quad}\eqno(2.13)$$ 
as 
${\,}r{\,}\rightarrow{\,}0_{+}{\,}$. 
To examine the boundary condition on the 
${\,}\xi_{k\ell}(r){\,}$ 
as 
${\,}r\rightarrow\infty{\,}$, 
note that the potential 
${\,}V_{\ell}(r)$ 
decreases sufficiently rapidly, as 
${\,}r\rightarrow\infty{\,}$, 
that a real solution to Eq.(2.12) behaves near 
$\{r=\infty\}$ 
according to
$$\xi_{k\ell}(r){\quad}
\sim{\quad}\Bigl(z_{k\ell}{\;}\exp\bigl(ikr_{S}^{*}\bigr){\,}
+{\;}z_{k\ell}^{*}{\;}\exp\bigl(-{\,}ikr_{S}^{*}\bigr)\Bigr)
{\quad}.\eqno(2.14)$$ 
Here, 
the 
$z_{k\ell}$ 
are certain dimensionless complex coefficients, 
which must be determined 
{\it via} 
the differential equation (2.12) together with the regularity conditions. 
Further, 
there is a natural normalisation of the basis 
${\,}\{\xi_{k\ell}(r)\}$ 
of radial wave-functions, 
as discussed in detail in [6].\par
\smallskip
\indent 
Given an appropriately normalised basis 
${\,}\{\xi_{k\ell}(r)\}{\,}$ 
of radial wave-functions on the final hypersurface 
$\Sigma_{F}\,$, 
one can expand out the interior linearised classical boundary-value solution 
near 
$\Sigma_{F}{\,}$ 
in the form
$$\phi^{(1)}{\quad}
={\quad}{{1}\over{r}}{\;}{\,}\sum_{\ell =0}^{\infty}{\;} 
\sum_{m=-\ell}^{\ell}{\;}\int_{-\infty}^{\infty}{\;}
dk{\;}{\;}a_{k\ell m}{\;}{\,}\xi_{k\ell}(t,r){\;} 
{{\sin(kt)}\over{\sin(kT)}}{\;}{\,}Y_{\ell m}(\Omega)
{\quad}.\eqno(2.15)$$ 
Here, 
the 
(nearly-) 
real quantities 
$\{a_{k\ell m}\}$ 
characterise the final data.\par
\smallskip
\indent 
An analogous description holds for fields of all the spins 
${{1}\over{2}}{\,},{\,}1$ 
and 
$2$ 
that have so far been checked [2,5,7]. 
When considering perturbative boundary data for a field of any spin, 
posed on 
$\Sigma_{F}{\,}$ 
in describing a final state resulting from black-hole evaporation, 
we denote by 
${\,}\{a_{sk\ell mP}\}{\,}$ 
a set of analogous `Fourier-like' coefficients, 
where 
$s$ 
gives the particle spin, 
$k$ 
the frequency, 
$(\ell,m)$ 
the angular quantum numbers, 
and 
${\,}P{\,}={\,}{\pm}{\,}1$ 
the parity 
(for $s{\,}{\neq}{\,}0{\,}$). 
For simplicity, 
we study in this Paper only bosonic perturbations, 
of spins 
$s{\,}={\,}0,1,2$, 
as treated in [2-4,6-8]. 
In each case, 
we found that the quantum amplitude or wave functional 
is of the semi-classical form, 
being given by
$$\Psi\Bigl[\{a_{sk\ell mP}\}{\,};{\,}T\Bigr]{\quad}
={\quad}N{\,}\exp\biggl(i{\,}S_{\rm class}^{(2)}
\Bigl[\{a_{sk\ell mP}\}{\,};{\,}T\Bigr]\biggr) 
{\quad},\eqno(2.16)$$
where the pre-factor 
$N$ 
depends only on 
$T{\,}$.  
Here, 
$S^{(2)}_{\rm class}$ 
denotes the 
(second-variation) 
action of the classical infilling solution, 
as a functional of the boundary data, 
corresponding to Eq.(1.2) for the 
${\,}s=0{\,}$ 
case.
For simplicity, 
we denote the collection of indices in 
$a_{sk\ell mP}{\,}$ 
by 
$j{\,}$. 
Further, 
we write 
$M_{I}{\,}$ 
for the total 
(time-independent) 
ADM 
(Arnowitt-Deser-Misner) 
mass of the `space-time', 
as measured at spatial infinity [27]. 
The ADM mass 
$M_{I}{\,}$, 
which is the limit at large radius of the variable mass 
${\,}m(t,r)$ 
of the Vaidya metric, 
is a functional of the final field configurations 
$\{a_{j}\}$ 
on 
$\Sigma_{F}{\,}$, 
since it depends on the full gravitational field 
which results from finding the classical solution 
of the complexified boundary-value problem. 
In Sec.6.1, 
we discuss this relationship between the total energy 
and the final field configurations, 
in the context of the normalisation of the probability density 
associated with the boundary data 
$\{a_{j}\}$ 
on 
$\Sigma_{F}{\,}$.\par
\smallskip
\indent 
As was found 
(for example) 
in the scalar case 
$s=0{\,}$ 
in [2-4,8], 
the classical action is dominated by contributions from frequencies 
$k$ 
with the values
$$k{\quad}
={\quad}k_{n}{\quad}
={\quad}{{n\pi}\over{T}}{\qquad}; 
{\qquad}{\qquad}{\qquad}n{\,}={\,}1{\,},2{\,},3{\,},{\;}\ldots 
{\quad}.\eqno(2.17)$$ 
It is also useful to define 
${\Delta}k_{j}{\,}$ 
to be the spacing between neighbouring 
${\,}k_{j}$-values:
$$\Delta k_{j}{\quad}
={\quad}{{\pi}\over{T}} 
{\quad}.\eqno(2.18)$$
\smallskip
\indent 
In an analogous way [2,7,25], 
for the corresponding Dirichlet problem with 
${\,}s{\,}={\,}1{\,},2{\,}$, 
and for 
${\,}s{\,}={\,}0{\,}$
(then neglecting the polarisation $P{\,}$), 
we found that
$$\eqalign{&S^{(2)}_{\rm class}\Bigl[\{a_{sk\ell mP}\}{\,};{\,}T\Bigr]\cr
&={\quad}{\;}{{1}\over{4\pi}}{\,}
\sum_{s=0,1,2}{\;}\sum^{\infty}_{\ell =s}{\;} 
\sum^{\ell}_{m=-\ell}{\;}\sum_{P=\pm}{\;}c_{s}{\;}
{{(\ell -s)!}\over{(\ell +s)!}}{\;}P
\int^{R_{\infty}}_{0}dr{\;}e^{(a-b)/2}{\;}\xi_{s\ell m}{\,}
\Bigl(\partial_{t}\xi^{*}_{s\ell mP}\Bigr)
{\Bigl\vert}_{\Sigma_{F}}\cr 
&{\quad}{\;}{\;}-{\,}{{1}\over{2}}{\,}M_{I}{\,}T
{\quad}.\cr}\eqno(2.19)$$ 
\noindent 
Equivalently,
$$\eqalign{&S^{(2)}_{\rm class}
\Bigl[\{a_{sk\ell mP}\}{\,};{\,}T\Bigr]\cr
&{\;}={\,}\sum_{s}{\,}\sum_{\ell mP}
(-1)^{s}{\;}c_{s}{\;}{{(\ell -s)!}\over{(\ell +s)!}}{\,}
\int^{\infty}_{0}dk{\;}{\,}k{\;}{\,}
{\vert}z_{sk\ell P}{\vert}^{2}{\;}
{\bigl\vert}a_{sk\ell mP}{\,}+(-1)^{s}{\,}P{\,}a_{s,-k\ell mP}
{\bigr\vert}^{2}{\,}
\cot(kT)\cr
&{\quad}{\;}{\;}-{\,}{{1}\over{2}}{\,}M_{I}{\,}T
{\quad}.\cr}\eqno(2.20)$$ 
\noindent 
The parity operator 
$P$, 
taking the values 
${\,}\pm{\,}1{\,}$, 
is defined more explicitly through its action on the coefficients 
$\{a_{sk\ell mP}\}{\,}$, 
according to:
$$a_{sk\ell mP}{\quad} 
={\quad}P{\,}(-1)^{m}{\;}a^{*}_{s,-k\ell,-m P}
{\quad}.\eqno(2.21)$$
\noindent 
The coefficients 
${\,}c_{s}{\;}{\;}(s{\,}={\,}0{\,},1{\,},2{\,}){\,}$ 
are given by
$$c_{0}{\quad}={\quad}2\pi{\qquad},
{\qquad}c_{1}{\quad}={\quad}{{1}\over{4}}{\qquad},
{\qquad}c_{2}{\quad}={\quad}{{1}\over{8}}
{\quad}.\eqno(2.22)$$ 
\noindent 
For higher bosonic spins 
${\,}s{\,}={\,}1{\,},2{\,}$ 
in the adiabatic approximation above, 
the functions 
$\{\xi_{s\ell mP}(t{\,},r)\}$ 
obey equations similar to the adiabatic version of Eqs.(2.4,5) for 
$s=0{\,}$, 
but with a potential 
$V_{s\ell P}$ 
which depends on 
$s{\,}$:
$$\Bigl(e^{(b-a)/2}{\;}\partial_{r}\Bigr)^{2}{\,}\xi_{s\ell mP}{\;}
-{\;}\bigl(\partial_{t}\bigr)^{2}{\,}\xi_{s\ell mP}{\;} 
+{\;}V_{s\ell P}{\;}{\,}\xi_{s\ell mP}{\quad} 
={\quad}0
{\quad}.\eqno(2.23)$$ 
\noindent 
The explicit forms of 
$V_{s\ell P}$ 
for 
$s=1$ 
and 
$2$ 
are given in [2,7]. 
The complex coefficients 
$\{z_{sk\ell P}\}$ 
relate to the boundary conditions at spatial infinity 
for the radial part of the functions 
$\{\xi_{s\ell mP}(t,r)\}{\,}$, 
as in Eqs.(2.12,14) above for the case 
$s=0{\,}$. 
Further details of the calculations 
for spins 1 and 2 are given in [2,7].\par
\smallskip
\indent 
Eqs.(2.16,20) can be interpreted as giving a 
`coordinate-representation' amplitude for each
(square-integrable) 
final-field configuration specified by
$\{a_{sk\ell mP}\}{\,}$, 
given that, 
on the initial hypersurface 
$\Sigma_{I}{\;}{\,}\{t=0\}{\,}$, 
the perturbations vanish, 
or equivalently that the metric and background matter 
are spherically symmetric.  
Hence, 
${\,}{\mid}\Psi[\{a_{sk\ell mP}\}{\,};{\,}T]{\mid}^{2}{\,}$ 
is the probability density for finding the field 
in a final configuration labelled by 
$\{a_{sk\ell mP}\}$ 
at time separation 
$T{\,}$, 
measured at spatial infinity.\par
\smallskip
\indent
Following [2,7,25], 
the classical action functional 
$S^{(2)}_{\rm class}{\,}$ 
is found to be a sum over individual `harmonics' labelled by 
$j{\,}$, 
which depend on the corresponding indices 
$\{sk_{j}\ell mP\}$ 
through the quantity 
${\mid}A_{j}{\mid}^{2}{\,}$, 
defined by
$${\bigl\vert}A_{j}{\bigr\vert}^{2}{\quad}
={\quad}2{\;}(-1)^{s}{\;}c_{s}{\;}{\,} 
{{(\ell -s)!}\over{(\ell +s)!}}{\;}{\;} 
{\bigl\vert}z_{j}{\bigr\vert}^{2}{\;}{\,}
\Bigl{\arrowvert}a_{j}{\,}+{\,}(-1)^{s}{\,}P{\,}a_{s,-k_{j}\ell m P}
\Bigl{\arrowvert}^{2} {\quad}.\eqno(2.24)$$
\noindent 
The coefficients 
$c_{s}{\,}$ 
for bosonic spins 
$s$ 
are given by Eq.(2.22). 
For 
$s=0{\,}$, 
the quantities 
$z_{j}{\,}$ 
are the complex numbers appearing in Eq.(2.14), 
which arise in solving the adiabatic radial mode equation (2.12); 
similarly for spins 
$s=1$ 
and 
$2$. 
This leads to the separated form of the quantum amplitude:
$$\Psi\Bigl[\{A_{j}\}{\,};{\,}T\Bigr]{\quad}
={\quad}{\hat N}{\;}{\,}e^{-{\,}i{{1}\over{2}}M_{I}T}{\;}{\,}
\prod_{j}{\,}\Psi(A_{j}{\,};{\,}T) 
{\quad},\eqno(2.25)$$ 
\noindent
where 
${\hat N}$ 
also depends only on 
$T{\,}$.\par
\smallskip
\indent 
Taking the classical action 
$S^{(2)}_{\rm class}$ 
in the form found in [2,25] for the scalar 
$s=0$ 
case 
(for example), 
one deduces that the wave functional for given scalar boundary data 
can be written as
$$\Psi\Bigl[\{A_{j}\}{\,};{\,}T\Bigr]{\quad}
={\quad}N{\;}{\,}e^{-{\,}i{{1}\over{2}}M_{I}T}{\;}{\,}
\prod_{j}{\;}{{1}\over{2i\sin(k_{j}T)}}{\;} 
\exp\biggl[{\,}{{i}\over{2}}{\,}\bigl(\Delta k_{j}\bigr){\,}k_{j}{\,}
{\mid}A_{j}{\mid}^{2}{\,}\cot(k_{j}T)\biggr] 
{\quad}.\eqno(2.26)$$
This will be related to the coherent-state description 
in the following Section 3.\par
\medskip

\noindent {\bf 3.  Coherent states}

\medskip
\indent 
The quantum amplitude (2.26) can usefully be re-written with the help 
of the Laguerre and Hermite polynomials [31], 
so as to demonstrate the relation with coherent states. 
First, 
we introduce the associated Laguerre polynomials 
$L^{(m-n)}_{n}(x){\,}$, 
defined by
$$L^{(m-n)}_{n}(x){\quad}
={\quad}\sum^{n}_{p=0}{\;}{m\choose n-p}{\;}{{(-x)^{p}}\over{p!}}
{\quad}\eqno(3.1)$$
for 
${\,}m{\,}\geq{\,}n{\,}\geq{\,}0{\,}$. 
The usual Laguerre polynomials 
${\,}L_{n}(x){\,}$ 
[31] are given by
$$L_{n}(x){\quad}
={\quad}L^{(0)}_{n}(x) 
{\quad}.\eqno(3.2)$$ 
The completeness relation for the set 
$\{L_{n}(x)\}$ 
reads:
$$\sum^{\infty}_{n=0}{\;}e^{-(x/2)}{\;}L_{n}(x){\;}
e^{-(y/2)}{\;}L_{n}(y){\quad} 
={\quad}\delta(x,y) 
{\quad}.\eqno(3.3)$$
\indent
Writing 
${\,}z=x+iy{\,}$, 
consider now the function 
$L_{n}\bigl({\mid}z{\mid}^{2}\bigr){\,}$, 
which appears in the above quantum amplitude,
re-expressed as in Eq.(3.5) below. 
For 
${\,}n>0{\,}$, 
this cannot be written as a product of two 
(decoupled) 
wave functions of 
$x$ 
and 
$y$ 
in an excited state, 
due to pair correlations [32].  
But, 
in terms of Hermite polynomials 
${\,}H_{p}(x){\,}$ 
[31], 
one can expand
$$L_{n}(x^{2}+y^{2}){\quad}
={\quad}{{(-1)^{n}}\over{2^{2n}{\,}n!}}{\;}{\;}
\sum^{n}_{p=0}{\;}{n\choose p}{\;}
H_{2p}(x){\;}H_{2n-2p}(y)
{\quad}.\eqno(3.4)$$
\indent 
With the help of the Laguerre and Hermite polynomials, 
the quantum amplitude (2.26) for our 
${\,}s=0{\,}$ 
scalar boundary-value problem, 
arising in black-hole evaporation,
can be re-written following Appendix A, 
in the form:
$$\eqalign{\Psi\Bigl[\{A_{j}\}{\,};T\Bigr]{\quad}
={\quad}{\hat N}{\;}e^{-{\,}i{{1}\over{2}}M_{I}T}{\;}&
\exp\Bigl(-\sum_{j}{\;}(\Delta k_{j}){\;} 
k_{j}{\,}{\mid}A_{j}{\mid}^{2}/2\Bigr)\quad\times\cr
&\times{\quad}\prod_{j}{\;}{\,}\sum^{\infty}_{n=0}{\;}
e^{-{\,}2iE_{n}T}{\;}L_{n}
\Bigl[{\,}k_{j}{\,}\bigl(\Delta k_{j}\bigr){\,}
{\mid}A_{j}{\mid}^{2}\Bigr] 
{\;},\cr}\eqno(3.5)$$
where
${\,}E_{n}
=\bigl(n+{{1}\over{2}}\bigr){\,}k_{j}{\,}$ 
is the quantum energy of the linear harmonic oscillator.  
Note also the dependence of the quantum amplitude on 
${\mid}A_{j}{\mid}{\;}{\,}$ 
-- it is spherically symmetric.\par
\smallskip
\indent 
There is a strong connection 
between the Schr\"odinger-picture wave functions
$$\Psi_{nj}(x_{j}{\,},T){\quad}
={\quad}{{N}\over{\pi}}{\;}e^{-{\,}(x_{j}/2)}{\;}
e^{-{\,}2iE_{n}T}{\;}L_{n}(x_{j})
{\quad},\eqno(3.6)$$
appearing in the wave-function Eq.(3.5), 
where 
${\,}x_{j}{\,}
={\,}k_{j}{\,}(\Delta k_{j}){\;}{\mid}A_{j}{\mid}^{2}{\;}$,
and the exact quantum solutions to the forced harmonic oscillator [33]. 
In this approach, 
one considers a 1-dimensional harmonic oscillator [33], 
with Hamiltonian
$$H{\quad}
={\quad}{{p^{2}}\over{2\mu}}{\;}
+{\;}{{1}\over{2}}{\,}\mu{\,}\omega^{2}q^{2}{\;}
+{\;}q{\,}F(t)
{\quad},\eqno(3.7)$$
where 
$F(t)$ 
denotes an external force, 
$\mu$ 
the oscillator mass and 
$\omega $ 
the oscillator frequency.  
Assume that 
$F(t)=0{\,}$ 
for 
$t<t_{0}{\,}$ 
and for 
$t>T{\,}$, 
so that the asymptotic states, 
at early and late times 
$t{\,}$, 
are free-oscillator states. 
One can calculate the amplitude 
$A_{km}$ 
to make a transition from the free-oscillator state 
${\mid}m>$ 
(with $m$ particles) 
at early times 
${\,}t<t_{0}{\,}$, 
to the free-oscillator state 
${\mid}k>$ 
at late times 
${\,}t>T{\,}$.  
Define the `Fourier transform' of the force:
$$\beta{\quad}
={\quad}\int^{T}_{t_0}dt{\;}{\,}F(t){\;}e^{-{\,}i{\omega}t}
{\quad},\eqno(3.8)$$ 
and set
$$z{\quad}
={\quad}{{{\mid}\beta{\mid}^{2}}\over{2\mu{\,}\omega}}
{\quad}.\eqno(3.9)$$
It has been shown [34-37], 
in the case 
${\,}m{\,}\geq{\,}k{\,}$, 
that
$$A_{km}{\quad}
={\quad}e^{i\lambda}{\;}e^{-{\,}(z/2)}{\;}
\biggl({{k!}\over{m!}}\biggr)^{{1}\over{2}}{\;} 
\biggl({{i\beta}\over{\sqrt{2\mu{\,}\omega}}}\biggr)^{m-k}{\;}
L_{k}^{(m-k)}(z)
{\quad},\eqno(3.10)$$
where 
$\lambda$ 
is a real phase.  
This expression also gives 
$A_{km}$ 
for 
${\,}m{\,}\leq{\,}k{\,}$, 
since 
$A_{km}=A_{mk}{\,}$ 
is symmetric.\par
\smallskip
\indent 
In the adiabatic limit, 
in which the force 
$F(t)$ 
changes extremely slowly, 
one has
$z{\;}{\ll}{\;}1{\,}$, 
and, 
from general considerations, 
a state which begins as 
${\mid}k>$ 
must end up in the same state 
${\mid}k>$ 
after the time-dependent force has been removed. 
From Eq.(3.10), 
one has
$$A_{kk}{\quad}
={\quad}e^{i\lambda}{\;}\exp\bigl(-{\,}z/2\bigr){\;}L_{k}(z)
{\quad}.\eqno(3.11)$$ 
The corresponding probability that there should be no change 
in the number of particles is
${\;}{\mid}A_{kk}{\mid}^{2}{\,}
={\,}\exp(-z){\;}[L_{k}(z)]^{2}{\;}$.
Apart from the introduction of mode labels 
$j$ 
denoting the `quantum numbers' 
$\{sk\ell mP\}\,$, 
together with a necessary re-interpretation for 
$z{\,}$,
these amplitudes are effectively the wave functions Eq.(3.5) 
derived from our boundary-value problem.\par
\smallskip
\indent 
It will be useful to give a brief derivation of Eq.(3.10) 
in the context of the coherent-state representation. 
Coherent states 
${\mid}\alpha >$ 
can be regarded as displaced vacuum states; 
that is, 
[13]
$${\mid}\alpha>{\quad}
={\quad}D(\alpha){\,}{\mid}0>
{\quad},\eqno(3.12)$$ 
where
$$D(\alpha){\quad}
={\quad}\exp\Bigl(\alpha{\,}a^{\dagger}{\,}
-{\,}\alpha^{*}a\Bigr)
{\quad}\eqno(3.13)$$
is a unitary displacement operator, 
obeying
$$D^{\dagger}(\alpha){\quad }
={\quad} D^{-1}(\alpha){\quad}
={\quad}D(-{\,}\alpha) 
{\quad},\eqno(3.14)$$
and where the states 
${\mid}\alpha>$ 
are eigenstates of the annihilation operator 
$a$ 
with complex eigenvalue 
$\alpha{\,}$.  
Among quantum states for the harmonic oscillator, 
they are the closest to classical states, 
in that they attain the minimum demanded by the uncertainty principle. 
Coherent states form an over-complete set, 
and are not orthogonal.  
In terms of the Fock-number eigenstates
$${\mid}n>{\quad}
={\quad}{{(a^{\dagger})^{n}}\over{\sqrt{n!}}}{\;}{\,}{\mid}0>
{\quad},\eqno(3.15)$$
one has [13]
$${\mid}\alpha >{\quad}
={\quad}\exp\Bigl(-{\,}{\mid}\alpha{\mid}^{2}/2\Bigr){\;}{\,}
\sum^{\infty}_{n=0}{\;}{\,}{{\alpha^{n}}\over{\sqrt{n!}}}{\;}{\,}
{\mid}n>
{\quad}.\eqno(3.16)$$
The coherent state labelled by 
${\,}{\alpha}=0{\,}$ 
is the ground state of the oscillator.  
If, 
for example, 
the system started in a vacuum state, 
the amplitude to find it subsequently in a coherent state 
${\mid}\alpha >$ is 
$$<0{\mid}\alpha>{\quad}
={\quad}<0{\mid}D(\alpha){\mid}0>{\quad}
={\quad}\exp\Bigl(-{\,}{\mid}\alpha{\mid}^{2}/2\Bigr) 
{\quad},\eqno(3.17)$$ 
up to a phase.\par
\smallskip
\indent
Proceeding towards a derivation of Eq.(3.10) by coherent-state methods, 
we note, 
from the properties of displacement operators, 
that
$$\eqalign{D(\xi){\mid}\alpha>{\quad}
&={\quad}D(\xi){\,}D(\alpha)\,{\mid}0>\cr
&={\quad}\exp\Bigl[\Bigl(\xi\alpha^{*}
-\xi^{*}\alpha\Bigr)/2\Bigr]{\;}
D(\xi +\alpha){\mid}0>\cr 
&={\quad}\exp\Bigl[\Bigl(\xi\alpha^{*}
-{\xi}^{*}\alpha\Bigr)/2\Bigr]{\;}
{\mid}{\xi}+{\alpha}>
{\quad}.\cr}\eqno(3.18)$$
\noindent 
For later reference, 
one can further show that
$$D^{+}(\gamma){\,}D(\mu){\,}D(\gamma){\quad}
={\quad}D(\mu){\;}
\exp\Bigl(\gamma^{*}{\mu}-{\gamma}{\,}\mu^{*}\Bigr)
{\;}.\eqno(3.19)$$
Using Eqs.(3.16,18), 
one then has
$$<m{\mid}D(\xi){\mid}\alpha>{\quad}
={\quad}{{1}\over{\sqrt{m!}}}{\;}{\,}(\xi +\alpha)^{m}{\;}{\,}
\exp\biggl[-{\,}{{1}\over{2}}\Bigl({\mid}\alpha{\mid}^{2}
+{\mid}\xi{\mid}^{2}+2{\,}{\xi}^{*}\alpha\Bigr)\biggr]
{\quad},\eqno(3.20)$$
and
$$<m{\mid}D(\xi){\mid}\alpha>{\quad}
={\quad}\exp\Bigl(-{\,}{\mid}\alpha{\mid}^{2}/2\Bigr){\;}
\sum^{\infty}_{n=0}{\;}{\,}
{{\alpha^{n}}\over{\sqrt{n!}}}{\;}{\,}
<m{\mid}D(\xi){\mid}n> 
{\quad}.\eqno(3.21)$$
On equating these, 
one finds that
$$(1+y)^{m}{\;}{\,}\exp\Bigl(-{\,}y{\,}{\mid}\xi{\mid}^{2}\Bigr){\quad}
={\quad}\exp\Bigl({\mid}\xi{\mid}^{2}/2\Bigr){\;}{\,}
\sum^{\infty}_{n=0}{\;}{\,}{\sqrt{{m!}
\over{n!}}}{\;}{\,}
\xi^{n-m}{\;}{\,}y^{n}{\;}{\,}<m{\mid}D(\xi){\mid}n>
{\;}.\eqno(3.22)$$ 
But, 
from the generating function for the associated Laguerre polynomials [31],
$$(1+y)^{m}{\;}{\,}e^{-{\,}yx}{\quad}
={\quad}\sum^{\infty}_{n=0}{\;}{\,}L^{(m-n)}_{n}(x){\;}{\,}
y^{n}{\quad},
{\qquad}{\mid}y{\mid}{\,}<{\,}1 
{\quad},\eqno(3.23)$$
one deduces that the matrix element between initial and final states is
$$<m{\mid}D(\xi){\mid}n>{\quad}
={\quad}\biggl({{n!}\over{m!}}\biggr)^{{1}\over{2}}{\;}{\,}
\xi^{m-n}{\;}{\,}\exp\Bigl(-{\,}{\mid}\xi{\mid}^{2}/2\Bigr){\;}{\,}
L^{(m-n)}_{n}\Bigl({\mid}\xi{\mid}^{2}\Bigr)
{\quad},\eqno(3.24)$$
which agrees with Eq.(3.10), 
up to an unimportant phase factor.\par

\medskip

\noindent {\bf 4. Generalised coherent states}

\medskip
\indent 
The amplitudes 
${\,}<m{\mid}D(\xi){\mid}n>{\,}$, 
as in Eq.(3.24), 
can also be interpreted in terms of generalised coherent states 
${\,}{\mid}n,\alpha>$ 
of the harmonic oscillator [35]. 
We define 
$${\mid}n,\alpha>{\quad}
={\quad}e^{-{\,}iE_{n}t}{\;}D\Bigl(\alpha(t)\Bigr){\mid}n>
{\quad}.\eqno(4.1)$$
\noindent 
Then, 
in the Fock representation, 
one has
$${\mid}n,\alpha >{\quad}
={\quad}\sum^{\infty}_{m=0}{\,}<m{\mid}D
\Bigl(\alpha(0)\Bigr){\mid}n>\,{\mid}m>
{\;}e^{-{\,}iE_{m}t}
{\quad}.\eqno(4.2)$$
\noindent 
As may be seen in Eq.(4.7) below, 
for generalised coherent states, 
the ground state 
${\,}(n=0)$ 
is a coherent state and not a vacuum state. 
Generalised coherent states are to the coherent states 
what the Fock states 
${\,}{\mid}n>$ 
are to the vacuum state; 
that is, 
excited coherent states.  
In addition, 
one has, 
where 
$I$ 
denotes the identity operator [35]:
$$\eqalignno{I{\quad}
&={\quad}{1\over\pi}{\,}\int{\,}d^{2}\alpha{\quad}
{\mid}n,\alpha>{\,}<n,\alpha{\mid}\quad,&(4.3)\cr
<n,\beta{\mid}n,\alpha>{\quad}
&={\quad}L_{n}\Bigl({\mid}\alpha -\beta{\mid}^{2}\Bigr){\;} 
\exp\biggl(\beta^{*}{\alpha}-{1\over 2}\Bigl({\mid}
\alpha{\mid}^{2}+{\mid}\beta{\mid}^{2}\Bigr)\biggr)\quad,&(4.4)\cr
<n,\beta{\mid}\psi>{\quad}
&={\quad}{{e^{-{\mid}\beta{\mid}^{2}/2}}\over\pi}
\int{\,}d^{2}\alpha{\quad} L_{n}
\Bigl({\mid}\alpha -\beta{\mid}^{2}\Bigr){\;}
e^{\beta^{*}\alpha}{\;}e^{-{\mid}\alpha{\mid}^{2}/2}
<n,\alpha{\mid}\psi>
{\,},&(4.5)\cr}$$
\noindent 
for an arbitrary state 
${\,}{\mid}\psi>{\,}$, 
with the definition:
$${\int}d^{2}\alpha{\quad}
\equiv{\quad}{\int}d\Bigl[{\rm Re}{\,}(\alpha)\Bigr]{\;} 
d\Bigl[{\rm Im}{\,}(\alpha)\Bigr]
{\quad}.\eqno(4.6)$$ 
\noindent 
Note that, 
from Eq.(4.4) with 
${\,}{\beta}{\,}={\,}0{\,}$, 
one has
$$<n,0{\mid}n,\alpha>{\quad}\equiv{\quad}<n{\mid}n,\alpha>{\quad}
={\quad}e^{-{\mid}\alpha{\mid}^{2}/2}{\;}
L_{n}\Bigl({\mid}\alpha{\mid}^{2}\Bigr)
{\quad},\eqno(4.7)$$
\noindent
again giving Eq.(3.11) up to a phase.  
The initial state could be seen not as a vacuum state, 
but as a Fock state, 
and the final state as a generalised coherent state.\par
\smallskip
\indent 
As shown by Hollenhorst [38], 
the amplitudes of Eq.(3.5) have yet a further interpretation: 
they are the matrix elements for the transition from the state 
${\mid}k>$ 
to the state 
${\mid}m>$ 
under the influence of a 
(linearised) 
gravitational wave, 
with the force 
$F(t)$ 
proportional to the Riemann curvature tensor component 
$R_{xtxt}(t){\,}$:
$$F(t){\quad}
={\quad}\mu{\;}\ell{\;}R_{xtxt}(t){\quad}
={\quad}-{1\over 2}{\;}\mu{\;}\ell{\;}
(\partial_{t})^{2}{\,}h^{TT}_{xx}
{\quad},\eqno(4.8)$$
\noindent 
where 
${\,}\ell{\,}$ 
is the distance between two particles along the 
$x$-axis, 
each of mass 
${1\over 2}\mu{\,}$, 
while 
${\,}h^{TT}_{xx}{\,}$ 
is the transverse-traceless gravitational-wave component 
of the metric [27] and 
$x$ 
is the change in separation of the masses.\par
\smallskip
\indent 
In the context of black-hole evaporation, 
one expects that the r\^ole of the force 
is played by the time-dependent background space-time{~} 
-- {~}which approximates a Vaidya space-time in the high-frequency limit 
at late times [2,6,20,21]. 
The above calculations indicate an explicit mathematical connection 
between the theory of forced harmonic oscillators and certain amplitudes 
relating to the dynamical evolution of black holes.\par
\smallskip
\indent 
An important point which we should mention is that, 
under the influence of a time-dependent force, 
an initial vacuum state transforms into a coherent state. 
Below, 
we discuss how, 
when one changes a phase parameter of the perturbations 
appearing in their frequencies 
(parametric amplification), 
an initial vacuum state transforms into a squeezed vacuum state. 
This phase is not an oscillator phase, 
but the small angle, 
$\delta{\,}$, 
through which the time 
${\,}T{\,}$ 
at infinity is rotated into the lower complex plane.\par

\medskip

\noindent {\bf 5. Squeezed-state formalism}

\medskip

\indent 
In this Section and in the following Sec.6, 
we shall see how,
by rotating the asymptotic Lorentzian time 
$T$ 
in the complex plane, 
and in the case of spherically-symmetric initial matter 
and gravitational fields,
one obtains a quantum-mechanical highly-squeezed-state interpretation 
for the final state in black-hole evaporation, 
in the limit of an infinitesimal rotation angle.\par
\smallskip
\indent 
Grishchuk and Sidorov [16,17] 
were the first to formulate particle creation 
in strong gravitational fields explicitly in terms of squeezed states, 
although the formalism does appear in Parker's original paper 
on cosmological particle production [15].  
In [16,17], 
it was shown that relic gravitons 
(as well as other perturbations), 
created from zero-point quantum fluctuations as the universe evolves, 
should now be in a strongly-squeezed state. 
Squeezing is just the quantum process 
corresponding to parametric amplification.\par
\smallskip
\indent 
Black-hole radiation 
in the squeezed-state representation was first discussed in [16,17].  
The `squeeze parameter' 
${\,}r_{j}$ 
(see below) 
was there related to the frequency 
$\omega_{j}$ 
and the black-hole mass 
$M$ 
through
$${\rm tanh}\bigl(r_{j}\bigr){\;}{\,}
={\;}{\,}\exp\Bigl(-{\,}4\pi M{\,}\omega_{j}\Bigr)
{\quad}.\eqno(5.1)$$
In this language, 
the vacuum quantum state in a black-hole space-time for each mode 
is a two-mode squeezed vacuum. 
However, 
our approach to squeezed states in black-hole evaporation is new, 
arising from a boundary-value problem 
involving two asymptotically-flat spacelike hypersurfaces,
together with Feynman's 
$+{\,}i\epsilon$ 
prescription [3].  
We now give a brief account of quantum-mechanical squeezed states.\par
\smallskip
\indent 
A general one-mode squeezed state 
(or squeezed coherent state)
is defined [14] as 
$${\mid}\alpha,z>{\quad}
={\quad}D(\gamma){\;}S(r,\phi){\,}{\mid}0>{\quad}
={\quad}D(\gamma){\;}S(z){\;}{\mid}0> 
{\quad}.\eqno(5.2)$$ 
Here, 
$D(\gamma)$ 
is the single-mode displacement operator, 
and we define
$$S(r,\phi){\quad}
\equiv{\quad}S(z){\quad} 
={\quad}\exp\biggl({{1}\over{2}}
\Bigl(za^{2}-{\,}z^{*} a^{\dagger 2}\Bigr)\biggr)
{\quad}\eqno(5.3)$$
in terms of annihilation and creation operators 
$a$ 
and 
$a^{\dagger}{\,}$, 
respectively; 
we also define
$$z{\quad}
={\quad}r{\,}e^{-{\,}2i\phi}
{\quad}.\eqno(5.4)$$ 
Here, 
${\,}S(r,\phi){\;}\equiv{\;}S(z){\,}$ 
gives the unitary squeezing operator for 
${\,}{\mid}\alpha,z>{\,}$, 
obeying
$$S^{\dagger}(z){\,}S(z){\quad}
={\quad}S(z){\,}S^{\dagger}(z){\quad}
={\quad}1
{\quad},\eqno(5.5)$$
with 
${\,}\gamma{\,}$ 
given by
$$\gamma{\quad}
={\quad}\alpha{\,}\cosh{\,}r{\;}
+{\;}\alpha^{*}{\;}e^{-{\,}2i\phi}{\,}\sinh{\,}r
{\quad}.\eqno(5.6)$$
The state Eq.(5.2) is a Gaussian wave-packet, 
displaced from the origin in position and momentum space.  
While the 
(real) 
squeezing parameter 
${\,}r{\;}{\,}$ 
(${\,}0{\,}\leq{\,}r{\,}<{\,}{\infty}{\,}$) 
determines the magnitude of the squeezing, 
the squeezing angle 
${\,}\phi{\;}{\,}$ 
(${\,}{\mid}\phi{\mid}{\,}<{\pi}/2{\,}$) 
gives the distribution of the squeezing between conjugate variables.  
The squeezed vacuum state occurs when
${\,}\alpha{\,}={\,}0{\,}{\;}$: 
$${\mid}z>{\quad}
\equiv{\quad}{\mid}0,z>{\quad} 
={\quad}S(z){\mid}0> 
{\quad}.\eqno(5.7)$$ 
The limit of high squeezing occurs when
${\,}r{\;}{\gg}{\;}1{\,}$, 
for which the state 
${\,}{\mid}z>$ 
is highly localised in momentum space.\par
\smallskip
\indent  
Consider now the amplitude
$$A{\quad}
={\quad}<\alpha,z{\mid}D(\mu){\mid}\alpha,z>{\quad} 
={\quad}<z{\mid}D^{\dagger}(\gamma){\;}D(\mu){\;}D(\gamma){\mid}z>
{\quad}.\eqno(5.8)$$ 
\noindent 
One can use Eq.(3.19) to show that
$$A{\quad}
={\quad}<z{\mid}D(\mu){\mid}z>{\;}
\exp\Bigl(2i{\,}{\rm Im}\bigl(\gamma^{*}\mu\bigr)\Bigr)
{\quad},\eqno(5.9)$$ 
\noindent 
and then use Eq.(5.6) and the relation 
${\,}\alpha{\,}={\,}{\mid}\alpha{\mid}{\;}e^{i\phi}{\,}$ 
to show that [31]
$${\mid}A{\mid}^{2}{\quad}
={\quad}e^{-{\,}{\mid}\gamma{\mid}^{2}}{\quad}
={\quad}\exp\biggl(-{\,}{\mid}\alpha{\mid}^{2}
\Bigl(\cosh{2r}{\,}+{\,}\sinh{2r}{\,}\cos{2({\theta}+\phi)}\Bigr)\biggr)
{\quad}.\eqno(5.10)$$
\indent 
Single-mode squeezed operators do not conserve momentum,
since they describe the creation of particle pairs with momentum 
$k{\,}$.  
Two-mode squeezed operators, 
however, 
describe the creation and annihilation of two particles 
(waves) 
with equal and opposite momenta.
A two-mode squeeze operator has the form [38]
$$S(r,\phi){\quad}
={\quad}\exp\Biggl(r\biggl(e^{-2i\phi}{\;}a_{+}{\;}a_{-}{\,}
-{\,}e^{2i\phi}{\;}a^{\dagger}_{+}{\;}a^{\dagger}_{-}{\,}\biggr)\Biggr)
{\quad},\eqno(5.11)$$
where 
${\,}a_{\pm}$
and 
${\,}a^{\dagger}_{\pm}$ 
are annihilation and creation operators for the two modes,
respectively.\par
\smallskip
\indent 
Consider two conjugate operators, 
${\hat p}{\,}$ 
and 
${\hat q}{\,}$, 
with variances
$\Delta{\hat p}$ 
and 
$\Delta{\hat q}{\,}$.  
In the squeezed-state formalism, 
one may construct states such that
$\Delta{\hat p}{\,}$ 
and 
$\Delta{\hat q}{\,}$ 
are equal, 
taking the minimum value possible.
The name `squeezed' refers to the fact that the variance 
of one variable in a conjugate pair can go below the minimum 
allowed by the uncertainty principle 
(the squeezed variable), 
while the variance of the conjugate variable 
can exceed the minimum value allowed 
(the superfluctuant variable) 
[14,40,41]. 
The superfluctuant variable is amplified by the squeezing process, 
and it becomes possible to observe this variable macroscopically; 
in contrast, 
the subfluctuant variable is squeezed and becomes unobservable.  
In particle production, 
whether by black holes or in cosmology, 
the number operator is a superfluctuant variable, 
while the phase is squeezed.\par

\medskip

\noindent {\bf 6. Analytic continuation and the large-squeezing limit}

\medskip
\indent 
We return to the quantum state 
described in the Schr\"odinger picture by Eq.(2.20), 
giving the wave-functional 
${\,}\Psi[\{A_{j}\}{\,};{\,}T]$ 
for perturbative bosonic field configurations on the final surface
${\,}\Sigma_{F}{\,}$, 
labelled by `coordinates' 
${\,}A_{j}{\,}\equiv{\,}A_{s\ell mP}{\;}$.
Again, 
$T$ 
denotes the (Lorentzian) time separation, 
measured at spatial infinity, 
between the initial surface
$\Sigma_{I}$ 
and the final surface 
$\Sigma_{F}{\,}$.  
For convenience, 
we repeat Eq.(2.26) and Eq.(3.5): 
$$\eqalign{\Psi\Bigl[&\{A_{j}\}{\,};{\,}T\Bigr]\cr
={\;}&{\hat N}{\;}
e^{-{\,}i{{1}\over{2}}M_{I}T}{\;}
\prod_{j}{\;}\Psi(A_{j}{\,};{\,}T)\cr
\equiv{\;}&{\hat N}{\;}e^{-{\,}i{{1}\over{2}}M_{I}T}{\;}
\prod_{j}{\;}{{1}\over{2i\sin(k_{j}T)}}{\;}
\exp\biggl({{i}\over{2}}{\,}\bigl({\Delta}k_{j}\bigr){\,}k_{j}{\,}
{\mid}A_{j}{\mid}^{2}{\,}\cot(k_{j}T)\biggr)\cr
={\;}&{\hat N}{\,}
e^{-{\,}i{{1}\over{2}}M_{I}T}{\;}
\exp\Bigl(-\sum_{j}{\,}({\Delta}k_{j}){\,}k_{j}{\,}
{\mid}A_{j}{\mid}^{2}/2\Bigr){\;} 
\prod_{j}{\,}\sum^{\infty}_{n=0}{\;}e^{-{\,}2iE_{n}T}{\;}
L_{n}\Bigl(k_{j}{\,}\bigl(\Delta k_{j}\bigr){\,} 
{\mid}A_{j}{\mid}^{2}\Bigr)
{\;}.\cr}\eqno(6.1)$$ 
\noindent 
We now define
$$\eqalign{\Phi\Bigl[\{A_{j}\}{\,};{\,}T\Bigr]{\quad} 
&={\quad}N{\;}
e^{-{\,}i{{1}\over{2}}M_{I}T}{\;} 
\prod_{j}{\;}2i\sin\bigl(k_{j}{\,}T\bigr){\;}
\Psi_{j}\bigl(A_{j}{\,};{\,}T\bigr)\cr
&\equiv{\quad}N{\;}
e^{-{\,}i{{1}\over{2}}M_{I}T}{\;}
\prod_{j}{\;}\exp\biggl({{i}\over{2}}{\,}
\bigl({\Delta}k_{j}\bigr){\,}k_{j}{\,}
{\bigl\vert}A_{j}{\bigr\vert}^{2}{\,}\cot(k_{j}T)\biggr)\cr
&={\quad}N{\;}\exp\biggl(i{\,}S^{(2)}_{\rm class}
\Bigl[\{A_{j}\}{\,};{\,}T\Bigr]\biggr)
{\quad},\cr}\eqno(6.2)$$
\noindent 
where 
$N$ 
is a 
$T$-dependent prefactor.\par
\smallskip
\indent 
As described in the Introduction, 
the classical boundary-value problem 
is expected to become well-posed when one rotates 
the asymptotic time-interval 
$T$ 
into the complex, 
taking
$$T{\quad} 
={\quad}{\mid}T{\mid}{\;}e^{-{\,}i\delta}{\qquad};
{\qquad}0{\,}<{\,}\delta{\,}\leq{\,}{{\pi}\over{2}}
{\quad}.\eqno(6.3)$$
\noindent 
By contrast, 
if one did not rotate 
$T{\,}$, 
so that 
$T$ 
remained real and positive, 
then the `sum' in Eq.(6.2) would diverge, 
due to the simple poles on the real-frequency axis at
$$k_{j}{\quad}
={\quad}\sigma_{n}{\quad}
={\quad}{{n\pi}\over{T}}{\qquad};
{\qquad}n{\;}={\;}1{\,},{\,}2,{\;}\ldots
{\quad},\eqno(6.4)$$ 
\noindent 
assuming that 
${\,}k_{j}{\,}{\mid}A_{j}{\mid}^{2}{\,}$ 
remains suitably non-zero near 
${\,}k_{j}{\,}={\,}\sigma_{n}{\,}$.  
Thus, 
the quantum amplitude 
cannot be computed simply by working with space-times 
of Lorentzian signature. 
As in the Introduction, 
we follow Feynman [1] in adopting a 
$+{\,}i\varepsilon{\,}$ 
prescription. 
That is, 
we carry out the quantum calculation above for
${\,}{\delta}>0{\,}$, 
and then, 
at the end of the calculation, 
take the limit 
${\,}{\delta}{\,}{\rightarrow}{\,}0_{+}{\,}$ 
of the quantum amplitude, 
to obtain the Lorentzian amplitude.\par
\smallskip
\indent 
Note also that we will not eventually take the limit of infinite 
${\mid}T{\mid}$. 
However, 
we do expect that, 
in practice, 
${\mid}T{\mid}$ 
will far exceed the dynamical collapse time-scale for the black hole, 
which is of order 
${\pi}M_{I}{\,}$ 
[27].  
We shall see that, 
in our problem, 
the replacement Eq.(6.3) of real 
$T$ 
by
$T{\;}={\;}{\mid}T{\mid}{\;}\exp(-{\,}i\delta),
{\;}{\,}0{\,}<{\,}\delta{\,}\leq{\,}{\pi}/2{\;}$, 
leads to a squeezed state, 
with a high degree of squeezing for small 
${\,}\delta{\,}$.  
For a general 
${\,}\delta{\,}>{\,}0{\,}$, 
one has, 
from Eqs.(6.2,3):
$$\eqalign{\Phi\Bigl[\{A_{j}\}{\,};{\,}&T\Bigr]\cr
={\quad}&\Phi\Bigl[\{A_{j}\}{\,};
{\,}{\mid}T{\mid}{\,},{\,}\delta\Bigr]\cr
={\quad}&N{\;}e^{-{\,}i{{1}\over{2}}M_{I}{\mid}T{\mid}\cos\delta}{\;}
e^{-{\,}{{1}\over{2}}M_{I}{\mid}T{\mid}\sin\delta}{\quad}\times\cr
&\times{\quad}\prod_{j}{\;}
\exp\biggl(-{\,}{1\over 2}{\,}\bigl(\Delta k_{j}\bigr){\,}k_{j}{\,}
{\mid}A_{j}{\mid}^{2}{\,}
\coth\Bigl(k_{j}{\,}{\mid}T{\mid}{\,}\sin\delta{\,}
-{\,}i{\,}\phi_{j}\bigl({\mid}T{\mid}{\,},{\,}\delta\bigr)\Bigr)\biggr)\cr
={\quad}&N{\;}e^{-{\,}{{1}\over{2}}iM_{I}{\mid}T{\mid}\cos\delta}{\;} 
e^{-{\,}{{1}\over{2}}M_{I}{\mid}T{\mid}\sin\delta}{\;} 
\prod_{j}{\;}\exp\biggl(-{\,}{1\over 2}
\Bigl(\Omega^{(R)}_{j}+{\,}i{\,}\Omega^{(I)}_{j}\Bigr) 
\bigl({\Delta}k_{j}\bigr){\,}k_{j}{\,}{\mid}A_{j}{\mid}^{2}\biggr)
{\;}.\cr}\eqno(6.5)$$
\noindent 
Here, 
we define
$$\eqalignno{\phi_{j}\Bigl({\mid}T{\mid}{\,},{\,}\delta\Bigr){\quad}
&={\quad}-{\,}k_{j}{\,}{\mid}T{\mid}{\,}\cos\delta
{\quad},&(6.6)\cr 
\Omega^{(R)}_{j}\Bigl({\mid}T{\mid}{\,},{\,}\delta\Bigr){\quad}
&={\quad}{{\sinh\bigl(2{\,}k_{j}{\,}{\mid}T{\mid}{\,}\sin\delta\bigr)} 
\over{2{\,}\Bigl(\cosh^{2}\bigl(k_{j}{\,}{\mid}T{\mid}{\,}\sin\delta\bigr)
-{\,}\cos^{2}\phi_{j}\Bigr)}}
{\quad},&(6.7)\cr
\Omega^{(I)}_{j}\Bigl({\mid}T{\mid}{\,},{\,}\delta\Bigr){\quad}
&={\quad}-{\,}{{\sin\bigl(2{\,}\phi_{j}\bigr)} 
\over{2{\,}\Bigl(\cosh^{2}\bigl(k_{j}{\,}{\mid}T{\mid}{\,}\sin\delta\bigr) 
-{\,}\cos^{2}\phi_{j}\Bigr)}}
{\quad}.&(6.8)\cr}$$
\noindent 
One can further re-write Eq.(6.5) in the form:
$$\eqalign{{\Phi}\Bigl[\{A_{j}\}&{\,};
{\,}{\mid}T{\mid}{\,},{\,}\delta\Bigr]\cr
=&{\;}N{\,}e^{-{\,}{{1}\over{2}}iM_{I}{\mid}T{\mid}\cos\delta}{\,}
e^{-{\,}{{1}\over{2}}M_{I}{\mid}T{\mid}\sin\delta}{\,}
\prod_{j}{\,}\exp\biggl(-{\,}{{1}\over{2}}{\,}
\bigl(\Delta k_{j}\bigr){\,}k_{j}
\biggl({1+e^{2i\phi_{j}}{\tanh}r_{j}
\over{1-e^{2i\phi_{j}}{\tanh}r_{j}}}\biggr)
{\,}{\bigl\vert}A_{j}{\bigr\vert}^{2}\biggr)
{\,},\cr}\eqno(6.9)$$ 
\noindent 
where we have set
$${\tanh r_{j}}\Bigl({\mid}T{\mid}{\,},{\,}\delta\Bigr){\quad}
={\quad}\exp\bigl(-{\,}2k_{j}{\,}{\mid}T{\mid}{\,}\sin\delta\bigr)
{\quad}.\eqno(6.10)$$
\noindent 
Therefore,
$$\exp\bigl(-{\,}2r_{j}\bigr){\quad}
={\quad}\tanh\bigl({\,}k_{j}{\,}{\mid}T{\mid}{\,}\sin\delta\bigr)
{\quad}.\eqno(6.11)$$
\noindent 
Hence, 
${\,}r_{j}\to 0$ 
for high frequencies, 
while 
${\,}r_{j}\to\infty$ 
for low frequencies.
On comparing with Sec.5, 
we recognise Eq.(6.9) as the coordinate-space representation 
of a quantum-mechanical squeezed state [23,42], 
with 
${\,}r_{j}({\mid}T{\mid}{\,},{\,}\delta)$ 
the squeeze parameter and
${\,}\phi_{j}({\mid}T{\mid}{\,},{\,}\delta)$ 
the squeeze angle.  
The evolution of the squeezed state is taken into account by the 
${\mid}T{\mid}$ 
dependence in 
${\,}r_{j}$ 
and in 
$\phi_{j}{\,}$,
which are in general complicated functions of time.  
Eq.(6.6) becomes simpler in the limit of infinitesimal 
$\delta{\,}$. 
Neglecting terms of 
$O(\delta^{2}){\,}$,
one has 
${\,}\phi_{j}({\mid}T{\mid}{\,},{\,}\delta){\;}
\simeq{\,}-{\,}k_{j}{\,}{\mid}T{\mid}{\,}$,
corresponding to free evolution.\par
\smallskip
\indent 
Computing the probability density
${\,}{\bigl\vert}\Phi[\{A_{j}\}{\,};
{\,}{\mid}T{\mid}{\,},{\,}\delta]{\bigr\vert}^{2}{\,}$, 
one finds that
$$\eqalign{{\Bigl\vert}{\Phi}[\{A_{j}\}&{\,};
{\,}{\mid}T{\mid}{\,},{\,}\delta{\,}]{\Bigl\vert}^{2}\cr
=&{\quad}{\bigr\vert}N{\bigl\vert}^{2}{\;}
\exp\Bigl(-{\,}M_{I}{\,}{\mid}T{\mid}{\,}\sin\delta\Bigr){\;}
\prod_{j}{\,}
\exp\biggl(-{\,}{{\coth\epsilon_{j}}
\over{f(k_{j}{\,},{\,}\epsilon_{j}{\,},{\,}{\mid}T{\mid}{\,})}}{\,}
\bigl({\Delta}k_{j}\bigr){\,}k_{j}{\,} 
{\bigl\vert}A_{j}{\bigr\vert}^{2}\biggr)
{\;},\cr}\eqno(6.12)$$
\noindent 
where
$$f\Bigl(k_{j}{\,},{\,}\epsilon_{j}{\,},
{\,}{\mid}T{\mid}\Bigr){\quad}
={\quad}1{\;}+{\;}{{\sin^{2}\bigl(k_{j}{\,}{\mid}T{\mid}\bigr)} 
\over{\sinh^{2}\epsilon_{j}}}
{\quad},\eqno(6.13)$$ 
\noindent 
and
$$\epsilon_{j}{\quad} 
={\quad}k_{j}{\,}{\mid}T{\mid}{\,}\sin\delta
{\quad}.\eqno(6.14)$$
\indent 
Eq.(6.12) describes a Gaussian non-stationary process, 
in that the variance is an oscillatory function of time.
One is now dealing with standing bosonic waves, 
rather than with travelling waves;  
the 
(classical) 
amplitudes for left- and right-moving waves are large and almost equal, 
much as in the cosmological scenario [16,17]. 
These standing waves imply a correlation between particles 
with opposite frequencies 
(and azimuthal angular momenta 
$\pm{\,}m{\,}$) 
in the final state. 
One consequence of the high-squeezing behaviour 
is that the variance for the amplitudes 
$\{A_{j}\}$ 
is large, 
so that there are large statistical deviations 
of the observable power spectrum from its expected value. 
This is just a manifestation of the Uncertainty Principle.\par 
\smallskip
\indent 
We now assume that 
$\delta$ 
is {~}sufficiently {~}small {~}that
${\,}\epsilon_{j}{\,}\ll{\,}1{\,},$ 
or, 
{~}equivalently, 
{~}that
$0{\,}<{\,}\delta{\,}\ll{\,}\bigl(k_{j}{\,}|T|\bigr)^{-1}{\,}$. 
Then, 
from Eq.(6.11):
$$\epsilon_{j}\quad
\simeq\quad\exp(-{\,}2r_{j}){\qquad},
{\qquad}\epsilon_{j}{\,}\ll{\,}1
{\quad},\eqno(6.15)$$ 
\noindent 
corresponding to 
${\,}r_{j}{\,}\gg{\,}1{\,}$, 
which is the high-squeezing limit. 
Thus, 
the high-squeezing limit corresponds to the limit 
${\,}\delta\to 0_{+}$ 
with 
${\,}k_{j}{\,}|T|{\,}$ 
bounded. 
A broadening of the width of the position distribution, 
which is of order 
$\bigl[\Omega_{j}^{(R)}\bigr]^{-{{1}\over{2}}}{\,}$,
therefore corresponds to a large squeezing in the momentum distribution.\par
\smallskip
\indent 
In the squeezed-state formalism, we regard this as the classical limit, 
since the average number of particles in the final state is large:
$$<N_{j}>{\quad} 
={\quad}4{\,}\sinh^{2}{\,}r_{j}{\quad} 
\simeq{\quad}\exp(2{\,}r_{j})
{\quad},\eqno(6.16)$$ 
for 
${\,}r_{j}{\,}\gg{\,}1{\,}$. 
For another way in which to view this, 
consider again the state Eq.(6.5).  
The WKB condition is met when
$$\Biggl\arrowvert{{\Omega^{(I)}_{j}}
\over{\Omega^{(R)}_{j}}}\Biggl\arrowvert{\quad} 
={\quad}\biggl\arrowvert{{\sin(2{\,}\phi_{j})} 
\over{\sinh\bigl(2{\,}k_{j}{\,}{\mid}T{\mid}{\,}\sin\delta\bigr)}}
\biggl\arrowvert{\quad}
\gg{\quad}1
{\quad},\eqno(6.17)$$
\noindent 
which is satisfied in the high-squeezing limit 
${\,}\epsilon_{j}{\,}\ll{\,}1{\,}$. 
The final state of the remnant black-hole evaporation flux, 
therefore, 
becomes more classical in the WKB {~}sense {~}in {~}the {~}limit 
$\delta{\,}\rightarrow{\,}0{\,}$.  
In this limit, 
one can effectively consider the final perturbations 
as being represented by a classical probability distribution function 
[16,17,40,43].  
As in the inflationary scenario in cosmology, 
the perturbations 
away from the spherically-symmetric black-hole background space-time, 
of quantum-mechanical origin, 
cannot be distinguished from classical stochastic perturbations, 
without the need of an environment for decoherence.\par
\smallskip
\indent 
An inflationary analogue 
can also be described for the initial conditions on the perturbations 
in the black-hole case. 
In cosmology, 
the assumption is that, 
at some early time just prior to inflation, 
the modes are in their adiabatic ground state.  
This, 
in turn, 
originates from the assumption that the universe 
was in a maximally-symmetric state at some time in the past [18].  
A similar assumption is present in our black-hole case, 
where we assumed that the initial perturbations were very weak,
so that the initial `star' and its gravitational field 
were spherically symmetric.\par
\smallskip
\indent 
To obtain one further view on the late-time state, 
for small 
$\epsilon_{j}{\,}$, 
one can express Eq.(6.12) in the form
$$\eqalign{{\Bigl\vert}
\Phi[\{A_{j}\}{\,};&{\,}{\mid}T{\mid}{\,},
{\,}{\delta}{\,}{\to}{\,}0{\,}]{\Bigl\vert}^{2}\cr
&={\quad}{\mid}N{\mid}^{2}{\;}
\exp\bigl(-{\,}M_{I}{\,}{\mid}T{\mid}{\,}\delta\bigr){\;} 
\prod_{j}{\,}\exp\biggl(-{\;}{{\epsilon_{j}}
\over{(\epsilon_{j})^{2}+{\,}\sin^{2}\bigl(k_{j}{\,}
{\mid}T{\mid}\bigr)}}{\,}\bigl({\Delta}k_{j}\bigr){\,}
k_{j}{\,}{\mid}A_{j}{\mid}^{2}\biggr)\cr
&\simeq{\quad}{\mid}N{\mid}^{2}{\;}
\exp\bigl(-{\,}M_{I}{\,}{\mid}T{\mid}{\,}\delta\bigr){\;}
\prod_{j}{\,}\exp\biggl(-{\,}\bigl({\Delta}k_{j}\bigr){\,}k_{j}{\,}
{\mid}A_{j}{\mid}^{2}{\;}\rho_{j}\biggr)
{\,},\cr}\eqno(6.18)$$
\noindent 
where we have defined
$$\rho_{j}{\quad}
={\quad}\sum^{\infty}_{n=-\infty}\bigl(\Delta\omega_{n}\bigr){\;}
\delta(k_{j}-{\,}\omega_{n}){\quad}
={\quad}\sum^{\infty}_{n=-\infty}{\,}
\exp\bigl(2in{\,}k_{j}{\,}|T|\bigr)
{\;},\eqno(6.19)$$
and used the delta-function identities
$$\delta(x){\quad} 
={\quad}{{1}\over{\pi}}{\;} \lim_{\epsilon\rightarrow 0}{\,}
{{\epsilon}\over{\bigl({\epsilon}^{2}{\,}+{\,}x^{2}\bigr)}}
{\quad},\eqno(6.20)$$
and
$$\delta\bigl(f(x)\bigr){\quad} 
={\quad}\sum_{i}{\;}{\delta(x-x_{i})\over{|f'(x_{i})|}}
{\quad},\eqno(6.21)$$
\noindent 
where the 
$x_{i}$ 
are defined to be the zeros of 
${\,}f(x){\,}$. 
In these equations, 
we are also using the definitions
${\;}\omega_{n}{\,}={\,}n\pi/{\mid}T{\mid}{\,}$ 
and
${\;}\Delta\omega_{n}{\;}
\equiv{\;}\omega_{n+1}-\omega_{n}{\;}
={\,}\pi/{\mid}T{\mid}{\;}$.
Hence, 
interchanging the sums over 
$j$ 
and 
$n{\,}$, 
in the continuum limit for the 
$\{k_{j}\}$ 
frequencies, 
one has
$${\Bigl\vert}\Phi\bigl[\{A_{j}\}{\,};
{\,}{\mid}T{\mid}{\,},{\,}\delta{\,}{\rightarrow}{\,}0{\,}\bigr]
{\Bigl\vert}^{2}{\quad} 
\sim{\quad}{\mid}N{\mid}^{2}
\prod_{s\ell mP}{\,} 
\prod^{n_{\rm {max}}}_{n=1}{\,}
\exp\Bigl(-{\,}\bigl(\Delta\omega_{n}\bigr){\,}\omega_{n}{\,}
{\mid}A_{sn\ell mP}{\mid}^{2}\Bigr)
{\quad},\eqno(6.22)$$
\noindent 
in the small-$\delta$ limit. 
The sum 
${\,}\sum^{\infty}_{n=-\infty}{\,}$ 
has been converted into 
${\,}\sum^{\infty}_{n=1}{\;}$,
noting that 
$k_{j}>0{\,}$, 
and also that
${\,}k_{j}{\,}{\mid}A_{j}{\mid}^{2}{\,}{\rightarrow}{\,}0{\,}$ 
as 
${\,}k_{j}{\,}{\rightarrow}{\,}0{\,}$, 
and we have introduced 
$n_{\rm max}{\;}$, 
the largest value of 
$n$ 
such that
${\,}\omega_{n_{\rm max}}{\,}={\,}M_{I}{\,}$, 
providing an effective cut-off in the product over 
$n{\,}$. 
This agrees with the result summarized in Eq.(19) of [3], 
where the calculation of the probability density 
was based on the contour-integration treatment of [2].\par
\smallskip
\indent 
The presence of the delta function in Eq.(6.18) indicates that, 
in the high-squeezing limit, 
the random variable 
${\,}\phi_{j}$ 
associated with the final state is squeezed to discrete values, 
independently of the quantum numbers 
$\{s\ell mP\}{\,}$.  
Note that it is only the squeeze phases 
$\{\phi_{j}\}$ 
of the 
(standing-wave) 
perturbations which are fixed and correlated in the high-squeezing limit.\par

\medskip 
\noindent {\bf 6.1 Normalisation}

\medskip
\indent 
We now discuss the normalisation factor 
${\,}{\mid}N{\mid}^{2}{\,}$ 
in the probability density Eq.(6.12). 
We consider the dimensionless variables 
$\{x_j\}{\,}$, 
as defined after Eq.(3.6).  
Now, 
${\,}{\mid}N{\mid}^{2}{\,}$
is determined by integrating the probability density Eq.(6.12) 
over the space of all 
$\{x_{j}\}{\,}$, 
since the sum of all probabilities of all possible configurations 
$\{x_{j}\}$ 
is unity. 
Hence, 
$$\eqalign{{\mid}N{\mid}^{2}{\quad}
&={\quad}\prod_{j}{\,}{{\cosh\epsilon_{j}{\,}\sinh\epsilon_{j}}
\over{\Bigl(\sinh^{2}\epsilon_{j}{\,}
+{\,}{\sin^{2}\bigl(k_{j}{\,}{\mid}T{\mid}\bigr)}\Bigr)}}\cr  
&={\quad}\prod_{j}{\,} 
{{1}\over{\Bigl(\cosh\bigl(2r_{j}\bigr){\,}
-{\,}\cos\bigl(2\phi_{j}\bigr){\,}\sinh\bigl(2r_{j}\bigr)\Bigr)}} 
{\quad}.\cr}\eqno(6.23)$$
\noindent 
One can verify from Eqs.(6.6,11) that this infinite product converges.\par
\smallskip
\indent 
There is, 
however, 
an ambiguity in the form of the normalisation factor 
due to the presence of the surface term,
${\,}-{\,}{{1}\over{2}}{\,}M_{I}{\,}T{\,}$, 
in the action Eq.(2.20), 
arising from the boundary at large radius 
joining the initial and final space-like hypersurfaces. 
The origin of this ambiguity lies in the fact that the total ADM mass 
$M_{I}$ 
is a functional of the final field configurations
$\{x_{j}\}{\;}$ 
-- {~}see Eq.(8.6) below.\par
\smallskip
\indent 
Thus, 
in this case,
$${\mid}N{\mid}^{2}{\quad}
={\quad}\prod_{j}{\,}
\Bigl(\lambda^{2}{\,}\Omega_{j}^{(R)}{\,}
+{\,}2{\,}\epsilon_{j}\Bigr)
{\quad},\eqno(6.24)$$
\noindent 
where we have re-introduced the factor of 
${\,}\lambda^{2}{\;}$: 
here, 
$\lambda$ 
is the small parameter used in the expansion 
of the metric and fields in Eq.(1.1). 
Ambiguities caused by surface terms in the action 
were also discussed in the squeezed-state formalism in [42].\par
\smallskip
\indent
One consequence of the high-squeezing behaviour is that the variance
for the amplitudes 
$\{A_{j}\}$ 
is large, 
so that there are large statistical deviations 
of the observable power spectrum from its expected value.  
This is just a manifestation of the uncertainty principle.  
Indeed, 
from Eq.(6.23) and the probability distribution Eq.(6.12), 
we find that the average value of 
$x_{j}$ 
has the form
$$<x_{j}>{\quad} 
={\quad}{{1}\over{\lambda^{2}}}{\;}
\Bigl(\cosh\bigl(2r_{j}\bigr)
-{\,}\cos\bigl(2\phi_{j}\bigr)\sinh\bigl(2r_{j}\bigr)\Bigr)
{\quad}.\eqno(6.25)$$ 
\indent 
Thus, 
in the large-squeezing limit 
${\,}r_{j}{\,}\gg{\,}1{\,}$, 
we have
$$<x_{j}>{\quad}
\simeq{\quad}{{1}\over{\lambda^{2}}}{\;}
\exp\bigl(2r_{j}\bigr){\,}\sin^{2}\bigl(\phi_{j}\bigr)
{\quad}.\eqno(6.26)$$ 
From Eqs.(6.14) and (6.15), 
one finds
$$<x_{j}>{\quad} 
\simeq{\quad}{{1}\over{\lambda^{2}}}{\;}
{{\sin^{2}\bigl(\phi_{j}\bigr)}\over{k_{j}{\,}|T|{\,}\delta}}
{\quad}.\eqno(6.27)$$ 
\indent
It is clear from this equation that 
$<x_{j}>$ 
is not a smooth function of 
$j{\,}$, 
since it has a large number of peaks,
and also has zeros when 
${\,}\phi_{j}{\,}={\,}n{\pi}{\,}$, 
for each integer 
$n{\,}$.
This is indicative of the standing-wave feature 
of our boundary-value problem.\par
\smallskip 
\indent 
For comparison, 
in inflationary cosmology the oscillation phases of standing waves 
have fixed values, 
giving rise to zeros in the power spectrum, 
which are characteristic of the CMBR.  
The power spectrum of cosmological perturbations 
in the present universe is not a smooth function of frequency.  
The standing-wave pattern, 
due to squeezing, 
induces oscillations in the power spectrum. 
This in turn produces Sakharov oscillations [43,44], 
due to metric and scalar perturbations, 
in the distribution of higher-order multipoles 
of the angular correlation function for the temperature anisotropies [24,45] 
in the CMBR, 
for all perturbations at a given time whose wavelength 
is comparable with or greater than the Hubble radius defined for that time. 
That is, 
the peaks and troughs of the angular power spectrum 
have a close relationship with the maxima and minima 
of the metric power spectrum. 
For long wavelengths, 
the power spectrum becomes smoother.\par
\medskip

\noindent {\bf 7. Entropy and Squeezing}

\medskip
\indent 
There have been many accounts of how to determine entropy generation 
in the squeezing formalism [32,39,42,43,46-49].  
Hu and Pavon [50] were the first to associate entropy generation 
with the monotonic increase in particle number with time, 
induced by parametric amplification in a vacuum cosmological space-time.
As squeezing is the quantum analogue of parametric amplification, 
one would expect that entropy production could be calculated 
{\it via} 
the squeezed-state formalism. 
This is indeed the case, 
although, 
as with any entropy calculation, 
the nature of the coarse-graining must be specified.  
For squeezing, 
this is particularly relevant since squeezed evolution is unitary:  
that is, 
there is in principle no loss of information in the evolution 
of the initial pure state to the final pure squeezed quantum state.  
The definition of entropy 
depends on how one chooses to measure the observables 
associated with the final squeezed states.  
For example, 
one can reduce the final density matrix 
with respect to a Fock or coherent-state basis [42], 
or use eigenstates of the superfluctuant variable.  
In [32,40,41], 
the loss of information comes from the increased dispersion 
of the superfluctuant operator.\par
\smallskip
\indent 
Following the work of [32,46-49], 
a universal form for the entropy density growth 
${\,}{\Delta}S_{j}{\,}$ 
holds for each mode, 
when one studies the classical limit of large average particle number, 
corresponding to the large-squeezing r\'egime, 
namely
$${\Delta}S_{j}{\quad} 
\simeq{\quad}2r_{j}{\qquad};
{\qquad}r_{j}{\,}\gg{\,}1
{\quad},\eqno(7.1)$$ 
\noindent 
irrespective of the particular coarse-graining.  
On calculating the von-Neumann entropy 
${\,}S$ 
from Eq.(6.12), 
using also Eq.(6.23), 
one finds that 
(in units such that $k_{B}=1$):
$$\eqalign{S{\quad} 
&={\quad}-{\,}\int{\,}\prod_{j}{\;}dx_{j}{\;}{\,}
P(x_{j}){\;}\log{P(x_{j})}\cr 
&={\quad}1{\;}+{\;}\sum_{j}{\;}
\log\Bigl(e^{2r_{j}}\sin^{2}\phi_{j}{\,} 
+{\,}e^{-{\,}2r_{j}}\cos^{2}\phi_{j}\Bigr)
{\quad}.\cr}\eqno(7.2)$$
\noindent 
Thus, 
the entropy (7.2) arises from our ignorance 
of the precise final radiation configuration.  
In the high-squeezing limit, 
one has, 
therefore,
$${\Delta}S{\quad} 
\simeq{\quad}2{\,}\sum_{j}{\;}r_{j}{\;}
+{\;}\sum_{j}{\;}\log\Bigl(\sin^{2}\phi_{j}\Bigr)
{\quad}.\eqno(7.3)$$ 
\noindent 
which agrees with Eq.(7.1) when 
${\,}r_{j}{\,}\gg{\,}1{\,}$, 
even if 
${\;}\sin\phi_{j}{\;}\simeq{\;}0{\;}$. 
Eq.(7.3) remains valid even if we take into account the ambiguity 
in the normalisation factor discussed in Sec.6.1.

\medskip

\noindent {\bf 8. Classical predictions}

\medskip
\indent 
We now discuss the way in which strong peaks 
in the wave function lead to some definite predictions.  
In quantum cosmology, 
wave functions 
$\Psi(q)$ 
are commonly peaked in such a way as to describe 
(semi-classically)
families of classical trajectories, 
corresponding to solutions of the Hamilton-Jacobi equation.  
Loosely speaking, 
such wave functions are peaked about correlations 
between coordinates and momenta.  
Such correlations may perhaps be described more clearly with the help 
of the Wigner function 
${\,}W(p,q){\,}$ [51]; 
in this reference, 
the correlations were precisely identified 
{\it via} 
${\,}W(p,q){\,}$.  
The opposite extreme 
-- no such correlation 
-- occurs when 
$W(p,q)$ 
factorises into a product of a function of position 
$q$ 
and a function of momentum 
$p{\,}$. 
On the other hand, 
when 
$W(p,q)$ 
is peaked about some `surface' in phase space, 
say
$\{p{\,}={\,}f(q)\}{\,}$, 
then the wave function predicts this particular correlation.\par
\smallskip
\indent 
An alternative proposal for measuring correlations was given in [52].  
There, 
classical correlations for a given Wigner function 
were predicted by means of projection onto coherent states, 
where position and momentum are equally (un)known, 
as in the classical theory. 
As demonstrated in [51], 
in the harmonic-oscillator case, 
the correlation between 
$p$ 
and 
$q$ 
is such that the Hamiltonian equals the classical energy.  
In the present paper, 
we arrive at similar conclusions.\par
\smallskip
\indent 
We now return to our wave function 
${\,}\Psi\bigl[\{A_{j}\}{\,};{\,}T\bigr]$ 
of Eqs.(2.25,26), 
describing the quantum amplitude 
for typical anisotropic perturbations of spins 
${\,}s{\,}={\,}0{\,},{\,}1$ 
and
$2$ 
about a final background spherically-symmetric gravitational 
and massless-scalar field 
$\bigl(\gamma_{\mu\nu}{\,},{\,}\Phi\bigr)$.  
This amplitude is further described in Eq.(3.5), 
in terms of a product over modes 
$j{\,}$, 
each term involving an exponential times a Laguerre polynomial.  
The relevant argument of each Laguerre polynomial 
is the dimensionless quantity 
$x_{j}{\,}$, 
given after Eq.(3.6). 
Following the above discussion in this Section, 
we look for predictions from the Heisenberg-picture wave-functional, 
given by a product over 
$j$ 
of the wave functions 
$$\Psi^{(H)}_{n_{j}}(x_{j}){\quad} 
={\quad}{{N}\over{\pi}}{\;}
\exp\bigl(-{\,}x_{j}/2\bigr){\;}L_{n_{j}}(x_{j})
{\quad}.\eqno(8.1)$$
\noindent 
If we were to restore all units, 
the dimensionless argument of the Laguerre polynomial 
would acquire a factor 
${\,}\hbar^{-1}{\,}$. 
In most cases of astrophysical interest, 
one then has 
${\,}x_{j}{\,}\gg{\,}1{\,}$.  
Examining the wave function (8.1) in the limit of large argument, 
note from [31] that
$$L_{n}(x){\quad} 
\sim{\quad}{{(-{\,}x)^{n}}\over{n!}}{\qquad},
{\qquad}{\rm as}{\quad}x{\;}\rightarrow{\;}\infty
{\quad}.\eqno(8.2)$$ 
\noindent 
For each 
$j{\,}$, 
one can find the peak in the wave function as a function of 
${\,}x_{j}{\,}$, 
which is at
$$n_{j}{\quad} 
={\quad}{{1}\over{2}}{\;}k_{j}{\,}\bigl(\Delta k_{j}\bigr){\,}
{\mid}A_{j}{\mid}^{2}{\;}\hbar^{-1}
{\quad}.\eqno(8.3)$$
\noindent 
Taking the spin-0 case, 
for example, 
this gives,
from Eqs.(2.24,8.3),
$$n_{j}{\quad} 
={\quad}2\pi{\,}k_{j}{\,}
\bigl({\Delta}k_{j}\bigr){\,}{\mid}z_{j}{\mid}^{2}{\;} 
{\Bigl{\arrowvert}}a_{j}{\,}
+{\,}a_{0,-k_{j}{\ell}mP}\Bigl\arrowvert^{2}
{\quad}.\eqno(8.4)$$
\noindent 
But in Sec.2 of [8] 
(see also [4]),
we showed in Eq.(2.17) that, 
for spin-0 perturbations, 
one has
$${\mid}b_{j}{\mid}^{2}{\;}{\quad}
={\quad}2\pi{\,}k_{j}{\;}{\mid}z_{j}{\mid}^{2}{\;}
{\Bigl\arrowvert}a_{j}{\;}
+{\;}a_{0,-k_{j}\ell mP}\Bigl\arrowvert^{2}
{\quad},\eqno(8.5)$$
\noindent 
where the 
$\{b{_j}\}$ 
are `Fourier amplitudes' associated with the radiation reaching 
${\,}{\cal I}^{+}$ 
(future null infinity).
Thus, 
Eq.(8.5) gives a match between the positive-frequency decomposition 
for particles reaching 
${\,}{\cal I}^{+}$
(travelling waves) 
and the boundary-value formulation employed in our papers [4,8], 
with final field configurations specified on the hypersurface
$\Sigma_{F}{\,}$, 
given by 
$\{t=T\}{\,}$ 
(standing waves). 
From Eqs.(8.4,5), 
one has 
${\,}n_{j}{\,}={\,}({\Delta}k_{j}){\,}{\mid}b_{j}{\mid}^{2}{\,}$, 
for each 
$j{\,}$. 
Thus, 
we find that
$$\sum_{j}{\,}\hbar{\,}n_{j}{\,}k_{j}{\quad}
={\quad}\sum_{j}{\;}({\Delta}k_{j}){\;}k_{j}{\;}
{\mid}b_{j}{\mid}^{2}{\quad} 
={\quad}M_{I}
{\quad}.\eqno(8.6)$$
\noindent 
The left-hand side is just the total energy of the radiated particles. 
The middle expression is the total energy 
in the massless-scalar fluctuations, 
which 
(in the absence of any gravitational radiation) 
equals the initial ADM mass 
${\,}M_{I}{\,}$.\par

\medskip

\noindent {\bf 9. Conclusion}

\medskip
\indent 
In this paper, 
we have illustrated many aspects of the boundary-value formulation 
for linearised integer-spin fields 
propagating in an evaporating black-hole space-time.  
For simplicity, 
we have taken here only the case in which one has initial fields 
which are spherically symmetric. 
When the 
(Lorentzian) 
proper-time separation 
${\,}T$ 
at spatial infinity between the initial and final hypersurfaces 
is deformed infinitesimally into the lower-half complex plane, 
following Feynman [1], 
one obtains a quantum-mechanical squeezed-state formalism.  
The large-squeezing limit is equivalent to the WKB limit, 
and corresponds to rotating 
${\,}T{\,}$ 
by only an infinitesimal angle 
${\,}\delta{\,}$ 
into the lower half-plane. 
Since the final highly-squeezed state is a pure state, 
the unpredictability 
associated with any final momentarily-naked singularity 
in the Lorentzian space-time is avoided.\par
\smallskip
\indent 
We found that, 
as in the cosmological case, 
the bosonic perturbations on the black-hole background 
can be regarded as a stochastic collection of standing waves, 
rather than as travelling waves, 
in the high-squeezing limit.  
This leads to the prediction of peaks in the power spectrum,
for `relic' radiation from an evaporating black hole, 
analogous to the Sakharov oscillations in the CMBR.\par
\smallskip
\noindent {\bf Acknowledgments}
\smallskip
\indent
We are grateful to the referees for constructive comments.\par
\medskip
\noindent {\bf Appendix A: Derivation of the wave function Eq.(3.5)}
\medskip
\indent 
Here, 
we describe the chain of relations leading, 
in the scalar case, 
from Eqs.(2.25,26) to Eq.(3.5), 
thus expressing the quantum amplitude 
$\Psi[\{A_{j}\}{\,};{\,}T]$ 
as an infinite product, 
over the index 
$j{\,}$, 
of suitable {~}exponentials {~}and {~}Laguerre {~}polynomials. 
We {~}first {~}set
$a{\,}={\,}\exp\bigl(-{\,}ik_{j}{\,}T\bigr){\;},
{\;}{\,}y{\,}={\,}0{\;}$ 
and 
${\;}x{\,}={\,}x_{j}{\,}$ 
in Mehler's formula [31]
$$\bigl(1-a^{2}\bigr)^{{1}\over{2}}{\;}
\sum^{\infty}_{p=0}{\,}e^{-{{1}\over{2}}(x^{2}+y^{2})}{\;}
{{a^{p}H_{p}(x)H_{p}(y)}\over{2^{p}{\,}p!}}{\quad} 
={\quad}e^{{{1}\over{2}}(x^{2}-y^{2})}{\,}
\exp\Bigl(-{\,}(x-ay)^{2}/(1-a^{2})\Bigr)
{\;},\eqno(A.1)$$ 
\noindent 
where
$$H_{p}(x){\quad}
={\quad}(-1)^{p}{\;}e^{x^{2}}{\,}
{{d^{p}}\over{dx^{p}}}\Bigl(e^{-x^{2}}\Bigr){\qquad}; 
{\qquad}p{\;}={\;}0{\;},1{\;},2{\;},{\ldots} 
{\quad}\eqno(A.2)$$ 
\noindent 
define the Hermite polynomials, 
which appeared already in Eq.(3.4). 
This normalisation of the 
$H_{p}(x)$ 
implies that
$${{1}\over{2^{p}{\,}p!{\,}\pi^{{1}\over{2}}}}{\;}
\int^{\infty}_{-\infty}{\;}dx{\;}{\,}\exp\bigl(-{\,}x^{2}\bigr){\,} 
\Bigl(H_{p}(x)\Bigr)^{2}{\quad} 
={\quad}1
{\quad}.\eqno(A.3)$$ 
\smallskip
\noindent 
With the above choice of 
${\,}a{\,}$, $y{\,}$ 
and 
${\,}x{\,}$, 
we find
$$\Bigl(2{\pi}i{\,}\sin\bigl(k_{j}T\bigr)\Bigr)^{-{{1}\over{2}}}{\;}
\exp\Biggl({{1}\over{2}}{\,}i{\,}(x_{j})^{2}
\cot\bigl(k_{j}T\bigr)\Biggr){\quad} 
={\quad}\sum^{\infty}_{p=0}{\;}\exp\Bigl(-{\,}iE_{p}{\,}T\Bigr){\;}{\,}
\psi_{p}(x_{j})\;\,\psi_{p}(0)
{\;},\eqno(A.4)$$ 
\noindent 
where 
${\,}E_{p}{\;}={\;}k_{j}{\,}\bigl(p+{{1}\over{2}}\bigr){\,}$, 
and we define
$$\psi_{p}(x_{j}){\quad} 
={\quad}{{\exp\Bigl(-{\,}{{1}\over{2}}{\,}(x_{j})^{2}\Bigr){\;}
H_{p}(x_{j})} 
\over{\Bigl(2^{p}{\,}p!{\,}\pi^{{1}\over{2}}\Bigr)^{{1}\over{2}}}}
{\quad},\eqno(A.5)$$ 
\noindent 
such that
$$\int^{\infty}_{-\infty}{\;}dx_{j}{\;}{\,}
{\bigl\vert}\psi_{p}(x_{j}){\bigl\vert}^{2}{\quad} 
={\quad}1
{\quad}.\eqno(A.6)$$ 
\noindent 
Only even terms contribute to the sum in Eq.(A.4), 
since [31]
$$\eqalignno{H_{2p}(0){\quad} 
&={\quad}{{\bigl(-1\bigr)^{p}{\,}\bigl(2p\bigr)!}\over{p!}}
{\quad},&(A.7)\cr
H_{2p+1}(0){\quad} 
&={\quad}0
{\quad}.&(A.8)\cr}$$ 
\noindent 
Hence,
$$\Bigl(2{\pi}i{\,}\sin(k_{j}T)\Bigr)^{-{{1}\over{2}}}{\;}
\exp\biggl({{1}\over{2}}{\,}i{\,}(x_{j})^{2}\cot(k_{j}T)\biggr){\quad}
={\quad}\sum^{\infty}_{p=0}{\;}
\exp\Bigl(-{\,}i{\,}E_{2p}{\,}T\Bigr){\;}
\psi_{2p}(x_{j}){\;}{\,}\psi_{2p}(0)
{\,}.\eqno(A.9)$$ 
\noindent 
In addition, 
we  now define, 
for each index 
$j{\,}$ 
in Eq.(3.5), 
suitably scaled versions of the real and imaginary parts 
of the complex quantity 
$A_{j}$ 
by:
$$x_{j}{\;}{\,}
={\;}{\,}\biggl({{k_{j}}\over{V}}\biggr)^{{1}\over{2}}{\;}
{\rm Re}(A_{j}){\qquad}, 
{\qquad}y_{j}{\;}{\,} 
={\;}{\,}\biggl({{k_{j}}\over{V}}\biggr)^{{1}\over{2}}{\;} 
{\rm Im}(A_{j})
{\quad}.\eqno(A.10)$$
Then, 
$$\eqalign{&{{\exp\Bigl({{1}\over{2}}ik_{j}{\,}V^{-1}{\,}
{\mid}A_{j}{\mid}^{2}\cot\bigl(k_{j}T\bigr)\Bigr)} 
\over{2{\pi}i{\,}\sin\bigl(k_{j}T\bigr)}}\cr
&{\quad}={\quad}\sum^{\infty}_{p'=0}{\;}\sum^{\infty}_{p=0}{\;}
\exp\Bigl(-{\,}i\bigl(E_{2p}+E_{2p'}\bigr)T\Bigr){\;}{\,}
\psi_{2p}(x_{j}){\;}{\,}\psi_{2p'}(y_{j}){\;}{\,}    
\psi_{2p}(0){\;}{\,}\psi_{2p'}(0)\cr
&{\quad}={\quad}\sum^{\infty}_{p=0}{\;}{\,}
\exp\bigl(-{\,}2iE_{p}{\,}T\bigr){\;}
\sum^{p}_{p'=0}{\;}\psi_{2p'}{\,}(x_{j}){\;}{\,}
\psi_{2p-2p'}{\,}(y_{j}){\;}{\,}\psi_{2p'}(0){\;}{\,}\psi_{2p-2p'}(0)\cr
&{\quad}={\quad}{\pi}^{-1}{\;}
\exp\Bigl(-{\,}k_{j}{\,}{\mid}A_{j}{\mid}^{2}/2V\Bigr)
{\quad}\times\cr
&{\qquad}{\quad}\times{\quad}
\sum^{\infty}_{p=0}{\;}{\,}{{(-1)^{p}{\;}
\exp\bigl(-{\,}2iE_{p}{\,}T\bigr)}\over{2^{2p}{\,}p!}}{\;}{\,}
\sum^{\infty}_{p'=0}{\;}{\,}{{p!}\over{(p')!{\;}(p-p')!}}{\;}{\,}
H_{2p'}(x_{j}){\;}{\,}H_{2p-2p'}(y_{j})
{\quad},\cr}\eqno(A.11)$$
\noindent 
where we define 
${\,}V^{-1}{\,}={\,}{\Delta}k_{j}{\;}$. 
In the limit
${\,}{\mid}T{\mid}\rightarrow 0_{+}{\,}$ 
(or rather,
${\,}k_{j}{\,}{\mid}T{\mid}{\,}\ll{\,}1{\,}$), 
since the eigenfunctions
${\,}\psi_{p}(x_{j}){\,}$ 
form a complete orthonormal set, 
one has
$$\eqalign{\lim_{k_{j}{\mid}T{\mid}{\rightarrow}0_{+}}
\Bigl(2{\pi}i{\,}\sin\bigl(k_{j}T\bigr)\Bigr)^{-1}{\;}
\exp\Bigl(i{\,}k_{j}{\,}{\mid}A_{j}{\mid}^{2}\cot(k_{j}T)/2V\Bigr)
{\quad} 
&={\quad}{\delta}(x_{j}){\,}{\delta}(y_{j})\cr 
&\equiv{\quad}
\delta^{(2)}\biggl(\biggl({{k_{j}}\over{V}}\biggr)^{{1}\over{2}}A_{j}\biggr)
{\;}.\cr}\eqno(A.12)$$
\noindent 
This suggests that 
${\,}A_{j}{\,}\rightarrow{\,}0{\;}$ 
as 
${\;}k_{j}{\,}{\mid}T{\mid}{\,}\rightarrow{\,}0_{+}{\;}$, 
agreeing with our initial condition that the fields 
are spherically-symmetric. 
We now use the identity Eq.(3.4) in Eq.(A.11). 
Taking the product over all 
$j{\,}$, 
and introducing a normalisation factor 
${\hat N}{\,}$, 
we find 
$$\eqalign {\Psi
\Bigl[\{A_{j}\}{\,};{\,}T\Bigr]{\;}  
&={\;}\prod_{j}{\,}\Psi(A_{j}{\,};{\,}T)\cr
&={\;}{\hat N}{\;}e^{-{\,}i{{1}\over{2}}M_{I}T}{\;}
\exp{\Bigl(-\sum_{j}k_{j}{\,}{\mid}A_{j}{\mid}^{2}/2V\Bigr)}{\,}
\prod_{j}{\;}\sum^{\infty}_{p=0}{\;}
e^{-{\,}2iE_{p}T}{\;}L_{p}\biggl({{k_{j}}\over{V}}{\;}
{\mid}A_{j}{\mid}^{2}\biggr)
{\,},\cr}\eqno(A.13)$$
\noindent 
which gives Eq.(3.5), 
where we have now included the contribution to 
$\Psi{\,}$ 
from the time-like boundary near spatial infinity. 
One can confirm that Eq.(A.13) also gives Eq.(2.26);  
this simply involves use of the generating function 
for Laguerre polynomials [31].\par

\parindent = 1 pt
\medskip
\noindent
{\bf References}
\medskip

\indent [1] Feynman, R P and Hibbs, A R 1965 
{\it Quantum Mechanics and Path Integrals} (New York: McGraw-Hill)\par 

\indent [2] Farley, A N St J 2002 
Quantum Amplitudes in Black-Hole Evaporation 
{\it Ph.D. Dissertation} University of Cambridge, UK;
Squeezed states in black-hole evaporation by analytic continuation 
({\it Preprint} gr-qc/0209113)\par 

\indent [3]  Farley, A N St J and D'Eath, P D 2004 
Scalar-field amplitudes in black-hole evaporation, 
{\it Phys Lett.} B {\bf 601} 184\par 

\indent [4] Farley, A N St J and D'Eath, P D 2005 
Bogoliubov transformations for amplitudes in black-hole evaporation 
{\it Phys Lett.} B {\bf 613} 181\par 

\indent [5] Farley, A N St J and D'Eath, P D 2005 
Spin-1/2 amplitudes in black-hole evaporation 
{\it Class. Quantum Grav.} {\bf 22} 3001\par 

\indent [6] Farley, A N St J and D'Eath, P D 2006 
Vaidya Space-Time in Black-Hole Evaporation 
{\it Gen. Relativ. Gravit.} {\bf 38} 425\par 

\indent [7] Farley, A N St J and  D'Eath, P D 2006 
Quantum amplitudes in black-hole evaporation: Spins 1 and 2 
{\it Ann. Phys. (N.Y.)} {\bf 321} 1334\par

\indent [8] Farley, A N St J and  D'Eath, P D 2006 
Bogoliubov transformations in black-hole evaporation 
{\it Int. J. Mod. Phys.} D, in press
({\it Preprint} gr-qc/0510043)\par

\indent [9] D'Eath, P D 1996 {\it Supersymmetric Quantum Cosmology} 
(Cambridge: Cambridge University Press)\par 

\indent [10] McLean, W  2000 
{\it Strongly Elliptic Systems and Boundary Integral Equations} 
(Cambridge: Cambridge University Press)\par

\indent [11]  Garabedian, P R 1964 {\it Partial Differential Equations}
(New York: Wiley)\par 

\indent [12] D'Eath, P D 1999 
Loop amplitudes in supergravity by canonical quantization 
{\it Fundamental Problems in Classical, Quantum and String Gravity} 
ed N S\'anchez (Paris: Observatoire de Paris) p 166
({\it Preprint} hep-th/9807028)\par

\indent [13] Glauber, R J 1963 
{\it Phys. Rev.} {\bf 131} 2766\par 

\indent [14] Schumacher, B L 1986 
{\it Phys. Rep.} {\bf 135} 317\par 

\indent [15] Parker, L 1969 
{\it Phys. Rev.} {\bf 183} 1057\par

\indent [16] Grishchuk, L P and Sidorov, Y V 1989 
{\it Class. Quantum Grav.} {\bf 6} L161\par 

\indent [17] Grishchuk, L P and Sidorov, Y V 1990 
{\it Phys. Rev.} D {\bf 42} 3413\par

\indent [18] Hartle, J B and Hawking, S W 1983 
{\it Phys. Rev.} D {\bf 28} 2960\par 

\indent [19] Halliwell, J S and Hawking, S W 1985 
{\it Phys. Rev.} D {\bf 31} 1777\par   

\indent [20] Vaidya, P C 1951 
{\it Proc. Indian Acad. Sci.} A {\bf 33} 264\par 

\indent [21] Lindquist, R W, Schwartz, R A and Misner, C W 1965 
{\it Phys Rev.} {\bf 137} 1364\par 

\indent [22] Page, D N 1976 
{\it Phys. Rev.} D {\bf 13} 198\par 

\indent [23] Kiefer, C 2001 
{\it Class. Quantum Grav.} {\bf 18} L151; 
2003  Is there an information-loss problem for black holes?
{\it Lect. Notes Phys.} {\bf 633} 84\par 

\indent [24] Grishchuk, L P 1996 
{\it Phys. Rev.} D {\bf 53} 6784\par 

\indent [25] Farley, A N St J and  D'Eath, P D 2006  
Coherent and squeezed states in black-hole evaporation 
{\it Phys. Lett.} B {\bf 634} 419\par

\indent [26] Jackson, J D 1975 {\it Classical Electrodynamics} 
(New York: Wiley)\par

\indent [27] Misner, C W, Thorne, K S  and Wheeler, J A 1973 
{\it Gravitation} (San Francisco: Freeman)\par 

\indent [28] Regge, T and Wheeler, J A 1957 
{\it Phys. Rev.} {\bf 108} 1063\par

\indent [29] Vishveshwara, C V 1970 
{\it Phys. Rev.} D {\bf 1} 2870\par 

\indent [30] Zerilli, F J 1970 
{\it Phys. Rev.} D {\bf 2} 2141\par 
 
\indent [31] Gradshteyn, I S and Ryzhik, I M 1965 
{\it Tables of Integrals, Series and Products}  
(New York: Academic Press)\par 

\indent [32] Gasperini, M and Giovannini, M  1993  
{\it Class. Quantum Grav.} {\bf 10} L133\par  

\indent [33] Schwinger, J 1953 {\it Phys. Rev.} {\bf 91} 728\par 

\indent [34] Roy, S M and Virendra Singh 1982 
{\it Phys. Rev.} D {\bf 25} 3413\par 

\indent [35] Satyanarayana, M V 1985 
{\it Phys. Rev.} D {\bf 32} 400\par 

\indent [36] Cahill, K E and Glauber, R J 1969 
{\it Phys. Rev.} {\bf 177} 1857\par

\indent [37] Cahill, K E and Glauber, R J 1969 
{\it Phys. Rev.} {\bf 177} 1882\par 

\indent [38] Hollenhorst, J N 1979 
{\it Phys. Rev.} D {\bf 19} 1669\par 

\indent [39] Hu, B L, Kang, G and Matacz, A L 1994 
{\it Int. J. Mod. Phys.} A {\bf 9} 991\par 

\indent [40] Polarski, D and Starobinskii, A 1996 
{\it Class.Quantum Grav.} {\bf 13} 377\par 

\indent [41] Casadio, R and Mersini, L  
Short distance signatures in cosmology: Why not in black holes? 
({\it Preprint} hep-th/0208050)\par 

\indent [42] Matacz, A L 1994 {\it Phys. Rev.} D {\bf 49} 788\par 
 
\indent [43] Albrecht, A, Ferreira, P, Joyce, M  and Prokopec, T
1994 {\it Phys. Rev.} D {\bf 50} 4807\par

\indent [44] Albrecht, A 
Coherence and Sakharov oscillations in the microwave sky, 
({\it Preprint} astro-ph/9612015)\par

\indent [45] Bose, S and Grishchuk, L P 2002 
{\it Phys. Rev.} D {\bf 66} 043529\par  

\indent [46] Prokopec, T 1993 
{\it Class. Quantum Grav.} {\bf 10} 2295\par 

\indent [47] Brandenberger, R, Mukhanov. V and Prokopec, T 1992
{\it Phys. Rev. Lett.} {\bf 69} 3603\par 

\indent [48] Gasperini, M and Giovannini, M 1993 
{\it Phys. Lett.} B {\bf 301} 334\par  

\indent [49] Kruczenski, M, Oxman, L E and Zaldarriaga, M 1994 
{\it Class. Quantum Grav.} {\bf 11} 2377\par 

\indent [50] Hu, B L and Pavon, D 1986 {\it Phys. Lett.} B {\bf 180} 329\par 

\indent [51] Halliwell, J J 1987 {\it Phys. Rev.} D {\bf 36} 3626\par 

\indent [52] Anderson, A 1990 {\it Phys. Rev.} D {\bf 42} 585\par

\end